\documentclass[12pt]{article}
\pdfoutput=1
\usepackage[nosort]{cite}
\usepackage{amsmath,amssymb}

\usepackage{color}

\usepackage{cite}
\usepackage{hyperref}
\usepackage{tikz}
\usepackage{slashed}
\usepackage{subcaption}

\usepackage{draft}

\def\({\left(}
\def\){\right)}

\begin{document}

\begin{titlepage}

\begin{center}

\hfill \\
\hfill \\
\vskip 1cm

\title{A Modified Villain Formulation of Fractons and Other Exotic Theories}

\author{Pranay Gorantla$^{1}$, Ho Tat Lam$^{1}$, Nathan Seiberg$^{2}$, and Shu-Heng Shao$^{2}$}

\address{${}^{1}$Physics Department, Princeton University, Princeton NJ, USA}
\address{${}^{2}$School of Natural Sciences, Institute for Advanced Study, Princeton NJ, USA}

\end{center}

\vspace{2.0cm}

\begin{abstract}

\noindent
We reformulate known exotic theories (including theories of fractons) on a Euclidean spacetime lattice.  We write them using the Villain approach and then we modify them to a convenient range of parameters.  The new lattice models are closer to the continuum limit than the original lattice versions.  In particular, they exhibit many of the recently found properties of the continuum theories including emergent global symmetries and surprising dualities.  Also, these new models provide a clear and rigorous formulation to the continuum models and their singularities.  In appendices, we use this approach to review well-studied lattice models and their continuum limits. These include the XY-model, the $\bZ_N$ clock-model, and various gauge theories in diverse dimensions.  This presentation clarifies the relation between the condensed-matter and the high-energy views of these systems.  It emphasizes the role of symmetries associated with the topology of field space, duality, and various anomalies.
\end{abstract}

\vfill

\end{titlepage}

\eject

\tableofcontents

\section{Introduction}\label{sec:intro}

The surprising discoveries of \cite{PhysRevLett.94.040402,PhysRevA.83.042330} have stimulated exciting work on fracton models.  This subject is reviewed nicely in \cite{Nandkishore:2018sel,Pretko:2020cko}, which include many references to the original papers.

One of the peculiarities of these models is that their low-energy behavior does not admit a standard continuum field theory description.  Finding such a description is important for two reasons.  First, it will give a simple universal framework to discuss fracton phases, will organize the distinct models, and will point to new models.  Second, since the field theory will inevitably be non-standard, this will teach us something new about quantum field theory.

\subsection{Overview of continuum field theories for  exotic models}

Following earlier work on such continuum field theories \cite{Slagle:2017wrc,Bulmash:2018knk,Shirley:2019wdf,Pretko:2018jbi,Slagle:2018swq,Slagle:2020ugk}, we initiated a systematic analysis of exotic field theories, including theories of fractons \cite{Seiberg:2019vrp, paper1, paper2, paper3, Gorantla:2020xap,Gorantla:2020jpy,Rudelius:2020kta}.
Our resulting theories are simple-looking, but subtle.  They capture the low-energy dynamics and the behavior of massive charged particles of the underlying lattice models as probe particles.

The main features of these exotic continuum field theories are the following:
\begin{enumerate}
\item Unlike the underlying lattice models, which are nonlinear, the low-energy continuum actions are quadratic, i.e., the theories are free.
\item The spatial derivatives in the continuum actions are such that we should consider discontinuous and even singular field configurations and gauge transformation parameters.  In fact, such discontinuities are essential in order to reproduce the microscopic lattice results.
\item Some observables, e.g., the ground state degeneracy and the spectrum of some charged states, are divergent in the continuum theory.  In order to make them finite, we need to introduce a UV cutoff, i.e., a nonzero lattice spacing $a$.  Even though these observables are divergent, the regularized versions are still meaningful.
\item Some of the continuum theories have emergent global symmetries, which are not present in the microscopic lattice models.  For example, winding symmetries and magnetic symmetries, which depend on continuity of the fields, are absent on the lattice, but are present in the low-energy, continuum theory.
\item Depending on the specific microscopic description, the global symmetry of the low-energy theory can involve a quotient of the global symmetry of the lattice model.  Some symmetry operators act trivially in the low-energy theory and we should quotient by them.
\item The analysis of the continuum theories leads to certain strange states that are charged under the original or the emergent symmetries with energy of order $1\over  a$.  Because of the singularities and the energy of these states, this analysis appears questionable and was referred to as an ``ambitious analysis.''
\item The continuum models exhibit surprising dualities between seemingly unrelated models.  These dualities are IR dualities, rather than  exact dualities, of the underlying lattice models.  They depend crucially on the precise global symmetries of the long-distance theories, including the emergent symmetries and the necessary quotients of the microscopic symmetry.  These dualities also map correctly the strange charged states we mentioned above.
\item The continuum models have peculiar robustness properties. (See \cite{paper1}, for a general discussion of robustness in condensed-matter physics and in high-energy physics.) Some symmetry violating operators, which could have destabilized the long-distance theory, have infinitely large dimension in that theory, and therefore they are infinitely irrelevant.  This comment applies both to some of the underlying symmetries of the microscopic models as well as to the emergent global symmetries.
\end{enumerate}

\subsection{Modified Villain lattice models}

The purpose of this paper is to explore further the lattice models, rather than their continuum limits.  We will deform the existing lattice models in a continuous way to find new lattice models with interesting properties.  In particular, despite being lattice models with nonzero lattice spacing $a$, they have many of the features of the continuum models we mentioned above.

Although this is not essential, we find it easier to use a discretized Euclidean spacetime lattice.  Then, following Villain \cite{Villain:1974ir}, we replace the lattice model with another model, which is close to it at weak coupling.  We replace the compact fields, which take values in $S^1$ or $\bZ_M$, by non-compact fields, which take values in $\bR$ and $\bZ$ respectively.  Then, we compactify the field space by gauging an appropriate $\bZ$ global symmetry.  In most cases, this is achieved by adding certain integer-valued gauge fields.

So far, this is merely the Villain version of the original model.  Then, we further modify the model by constraining the field strength of the new integer-valued gauge fields to zero.  We refer to this model as the \textit{modified Villain version} of the system.
The modified Villain versions of the ordinary XY model and the  $U(1)$ gauge theory have been previously constructed in \cite{Sulejmanpasic:2019ytl}.\footnote{We thank Z.\ Komargodski and T.\ Sulejmanpasic for pointing out this reference and related papers to us.}

Let us demonstrate this in the standard 2d Euclidean XY-model.  (See Appendix \ref{2dXY}, for a more detailed discussion of this model.)  The degrees of freedom are circle-valued fields $\phi$ on the sites of the lattice and the standard lattice action is
\ie\label{XY-actioni}
 \beta \sum_\text{link} [1- \cos(\Delta_\mu \phi)]~,
\fe
where $\mu=x,y$ labels the directions and $\Delta_\mu \phi$ are the lattice derivatives.
The standard Villain version of this action is
\ie\label{XY-Villain-actioni}
 \frac{\beta}{2} \sum_\text{link} (\Delta_\mu \phi - 2\pi n_\mu)^2~.
\fe
Here $\phi$ is a real-valued field and $n_\mu$ is an integer-valued field on the links.  This theory has the $\bZ$ gauge symmetry
\ie\label{XY-Villain-gaugesymi}
\phi \sim \phi + 2\pi k~, \qquad n_\mu \sim n_\mu + \Delta_\mu k~,
\fe
where $k$ is an integer-valued gauge parameter on the sites.  Next, we deform the model further by constraining the gauge invariant field strength of the gauge field $n_\mu$,
\ie
{\cal N}\equiv\Delta_x n_y-\Delta_y n_x~,
\fe
to zero \cite{Gross:1990ub}.  We will refer to this and similar constraints as flatness constraints.  We do that by adding a Lagrange multiplier $\tilde \phi$, and then the full action becomes
\ie\label{XY-modifiedVillain-actioni}
\frac{\beta}{2} \sum_\text{link} (\Delta_\mu \phi - 2\pi n_\mu)^2 + i \sum_\text{plaquette} \tilde \phi {\cal N}~.
\fe

We refer to the action \eqref{XY-modifiedVillain-actioni} as the modified Villain version of the original action \eqref{XY-actioni}.  We will analyze it in detail in Appendix \ref{2dXY}.

In the bulk of the paper, we will apply this procedure to the lattice models of \cite{PhysRevB.66.054526,Xu2008,Vijay:2016phm,plaqising,Bulmash:2018lid,Ma:2018nhd,
Radicevic:2019vyb,paper1,paper2,paper3}. These include, in particular, the X-cube model \cite{Vijay:2016phm}.
The resulting lattice models turn out to share some of the nice features of our continuum theories, even though they are on the lattice.  Comparing with the list above, these lattice models have the following features:
\begin{enumerate}
\item The actions are quadratic in the fields; these theories are free.
\item The fields and the gauge parameters are discontinuous on the lattice.  As we take the continuum limit, they become more continuous.  But some discontinuities remain.  In fact, our rules in \cite{Seiberg:2019vrp, paper1, paper2, paper3, Gorantla:2020xap,Gorantla:2020jpy,Rudelius:2020kta} about the allowed singularities in the fields and the gauge transformation parameters follow naturally from this lattice model.
\item Since these are lattice models, there is no need to introduce another regularization.
\item All the emergent symmetries of the continuum theories (except continuous translations) are exact symmetries of these lattice models. Starting with these models, there are no emergent symmetries.
\item These lattice models do not exhibit additional symmetries beyond those of the continuum models.  No quotient of the microscopic global symmetry is necessary.
\item The strange charged states with energy of order $1\over  a$ of the ``ambitious analysis'' of the continuum theories are present in the new lattice models and they have precisely the expected properties.
\item All the surprising dualities of the continuum models are present already on the lattice.  These are not IR dualities, but exact dualities. All of them follow from using the Poisson resummation formula
    \ie\label{Possonresummationi}
    &\sum_n \exp\left[-\frac{\beta}{2} (\theta-2\pi n)^2 + i n \tilde \theta\right] \\
    &\qquad \qquad= \frac{1}{\sqrt{2\pi \beta}} \sum_{\tilde n} \exp\left[-\frac{1}{2(2\pi)^2\beta} (\tilde \theta-2\pi \tilde n)^2 - \frac{i\theta}{2\pi}(2\pi \tilde n-\tilde \theta)\right]~.
    \fe
\item Our new lattice models have the same global symmetry as the low-energy continuum limit.  Therefore, there is no need to discuss the robustness of the low-energy theory with respect the operators violating these symmetries.  The analysis of robustness with respect to symmetry-violating operators should be performed in the low-energy continuum theory and it is the same in the original models and in these new ones.  We note that our lattice theory is natural once this new symmetry is imposed.  (See \cite{paper1} for a discussion of naturalness and its relation to robustness.)
\end{enumerate}

To summarize, we deform the original lattice models to their modified Villain versions.  The new models exhibit some of the special properties of the continuum theories even without taking the continuum limit.

Furthermore, it is clear that, at least for some range of coupling constants, the previous models and the new deformed models flow to the same long-distance theories, which are described by the continuum field theories mentioned earlier.

One interesting aspect of our new lattice models is that they exhibit global symmetries with 't Hooft anomalies.  For example, the model \eqref{XY-modifiedVillain-actioni} has a global $U(1)$ momentum symmetry and a global $U(1)$ winding symmetry.  These symmetries act locally (``on-site''), but they still have a mixed anomaly.  The anomaly arises because the Lagrangian density and even its exponential are not invariant under these two symmetries --- instead, only the action, or its exponential, is invariant. Conversely, if a global symmetry acts on-site and the Lagrangian density is invariant, it is clear that the symmetry can be gauged and there is no anomaly.  See Appendix \ref{2dXY}, for a more detailed discussion.

We should add another clarifying comment.  The original lattice model can have several different phases.   The Villain version of that model has the same phases.  However, this is typically not the case for the modified model.  In some cases it describes one of the phases of the original model and other phases that that model does not have.

For example, as we will discuss in detail in Appendix \ref{2dXY}, the model \eqref{XY-modifiedVillain-actioni} describes the large $\beta $ gapless phase of the 2d XY-model \eqref{XY-actioni} or \eqref{XY-Villain-actioni}.  But instead of describing its gapped phase with small $\beta$, it describes other continuum theories there.  This behavior is the same as that of the $c=1$ conformal field theory with arbitrary radius.

Another example, which we will discuss in Appendix \ref{AppeUone}, is the 3d $U(1)$ gauge theory.  The standard lattice model and its Villain version have a gapped confining phase \cite{POLYAKOV1977429}.  Our modified version of that model is gapless and is similar to the corresponding continuum gauge theory.

As we said above, some of our lattice models have global continuous symmetries with 't Hooft anomalies.  This means that their long-distance behavior must be gapless.  This is consistent with the fact that they are gapless even when the original lattice model is gapped.

Another perspective on these new lattice models is the following.  Since our exotic continuum models involve discontinuous field configurations, their analysis can be subtle.  The new lattice models can be viewed as rigorous presentations of the continuum models.  In fact, as we said above, they lead to the same answers as our continuum analysis including the more subtle ``ambitious analysis,'' thus completely justifying it.

In order to demonstrate our approach, we will use it in Appendices \ref{sec:VillainQM}, \ref{Villainlattia}, and \ref{pformgau} to review some well-known models.  In particular, we will present lattice models of various spin systems (including the XY-model \eqref{XY-actioni}) and gauge theories, which share many of the properties of their continuum counterparts.  In addition to demonstrating our approach, some people might find that discussion helpful.  It relates the condensed-matter perspective to the high-energy perspective of these theories.

\subsection{Outline}

Following \cite{paper1,paper2,paper3,Gorantla:2020xap,Gorantla:2020jpy}, Sections \ref{sec:2+1d} and \ref{sec:3+1d-cubic} are divided into three parts. We study an XY-type model, then the $U(1)$ gauge theory associated with the momentum symmetry of this XY-type model, and then the corresponding $\mathbb Z_N$ gauge theory. We present the modified Villain lattice action of each model, dualize it (if possible) using the Poisson resummation formula \eqref{Possonresummationi} for the integer-valued gauge fields, discuss the global symmetries, and take the continuum limit.  All these modified Villain lattice models exhibit all the peculiarities of the corresponding continuum theories of \cite{paper1,paper2,paper3}.

Even though we do not present it here, we have performed the same analysis for the exotic 3+1d continuum theories of \cite{Gorantla:2020jpy}, and we found similar results for the dualities and global symmetries of these modified Villain models. In particular, we have shown that the modified Villain formulation of the $\mathbb Z_2$ checkerboard model \cite{Vijay:2016phm} is exactly equivalent to two copies of the modified Villain formulation of the $\mathbb Z_2$ X-cube model.
 This equivalence can be regarded as the universal low-energy limit of the equivalence shown in the Hamiltonian formulation in \cite{Shirley:2019xnp}.

In Section \ref{sec:2+1d}, we study the modified Villain formulation of the exotic 2+1d continuum theories of \cite{paper1}.  These include systems with global $U(1)$ subsystem symmetry and $U(1)$ and ${\mathbb Z}_N$ tensor gauge theories.  We start with the XY-plaquette model of \cite{PhysRevB.66.054526} on a 2+1d Euclidean lattice, and present its modified Villain action.  Next, we study the modified Villain formulation of the associated $U(1)$ lattice tensor gauge theory. Finally, we present two equivalent $BF$-type actions of the $\mathbb Z_N$ lattice gauge theory: one with only integer fields (integer $BF$-action), and another with real and integer fields (real $BF$-action). All these modified Villain lattice models behave exactly as the corresponding continuum theories of \cite{paper1}.

In Section \ref{sec:3+1d-cubic}, we study the modified Villain formulation of the exotic 3+1d continuum theories of \cite{paper2,paper3}. Again, these include systems with global $U(1)$ subsystem symmetry and $U(1)$ and ${\mathbb Z}_N$ tensor gauge theories.  We present the modified Villain actions of the XY-plaquette model on a 3+1d Euclidean lattice, its associated $U(1)$ lattice tensor gauge theory, and the $\mathbb Z_N$ X-cube model.  As in Section  \ref{sec:2+1d}, these modified Villain models exhibit the same properties as their continuum counterparts in \cite{paper3}.

In three appendices we use our modified Villain formulation to review the properties of well-studied models. Some readers might find it helpful to read the appendices before reading Sections \ref{sec:2+1d} and \ref{sec:3+1d-cubic}.

Appendix \ref{sec:VillainQM} is devoted to some classic quantum-mechanical systems. We start with the problem of particle on a ring with a $\theta$-parameter. For $\theta\in \pi\bZ$, our Euclidean lattice model exhibits a mixed 't Hooft anomaly between its charge conjugation symmetry and its $U(1)$ shift symmetry.  We also use our Euclidean lattice formulation to study the quantum mechanics of a system whose phase space is a two-dimensional torus, a.k.a. the non-commutative torus.

In Appendix \ref{Villainlattia}, we discuss some famous 2d Euclidean lattice models using our modified Villain formulation. First, we study the modified Villain version of the 2d Euclidean XY-model.  Unlike the standard XY-model, it has an exact winding symmetry and an exact T-duality. It is very similar to the continuum $c=1$ conformal field theory of a compact boson.  Then, we study the 2d Euclidean $\mathbb Z_N$ clock-model by embedding it into the XY-model.

In Appendix \ref{pformgau}, we study $p$-form $U(1)$ gauge theories on a $d$-dimensional Euclidean spacetime lattice. We discuss their duality and the role of the Polyakov mechanism for $p=d-2$. We also study the $p$-form $\mathbb Z_N$ gauge theory. We briefly comment on the relation between $\mathbb Z_N$ toric code and the ordinary $\mathbb Z_N$ gauge theory.

\section{2+1d (3d Euclidean) exotic theories}\label{sec:2+1d}
In this section, we describe modified Villain lattice models corresponding to the exotic 2+1d continuum theories of \cite{paper1}. All lattice models discussed here are placed on a 3d Euclidean lattice with lattice spacing $a$, and $L^\mu$ sites in $\mu$ direction. We use integers $\hat x^\mu$ to label the sites along the $\mu$ direction, so that $\hat x^\mu \sim \hat x^\mu + L^\mu$.

Since the spatial lattice has a $\mathbb{Z}_4$ rotation symmetry, we will organize the fields according to the irreducible, one-dimensional representations $\mathbf{1}_n$ of $\mathbb{Z}_4$ with $n=0,\pm 1,2$ labeling the spin. In the discussion below, a field without any spatial index is in $\mathbf{1}_0$ and a field with  the spatial indices $xy$ is in $\mathbf{1}_2$.

\subsection{$\phi$-theory (XY-plaquette model)}
We start with a Euclidean spacetime version of the XY-plaquette model of \cite{PhysRevB.66.054526}.  The degrees of freedom are phases $e^{i\phi}$ at every site with the action
\ie\label{XYplaq-action}
\beta_0 \sum_{\tau\text{-link}} [1-\cos(\Delta_\tau \phi)] + \beta \sum_{xy\text{-plaq}} [1-\cos(\Delta_x \Delta_y \phi)]~.
\fe
At large $\beta_0,\beta$, we can approximate the action by the Villain action
\ie\label{XYplaq-Vill-action}
\frac{\beta_0}{2} \sum_{\tau\text{-link}} (\Delta_\tau \phi - 2\pi n_\tau)^2 + \frac{\beta}{2} \sum_{xy\text{-plaq}} (\Delta_x \Delta_y \phi - 2\pi n_{xy})^2~,
\fe
with real-valued $\phi$ and integer-valued $n_\tau$ and $n_{xy}$ fields on the $\tau$-links and the $xy$-plaquettes, respectively. We interpret $(n_\tau,n_{xy})$ as $\mathbb Z$ tensor gauge fields that make $\phi$ compact because of the gauge symmetry
\ie\label{XYplaq-Vill-gaugesym}
&\phi \sim \phi + 2\pi k~,
\\
&n_\tau \sim n_\tau + \Delta_\tau k~,
\\
&n_{xy} \sim n_{xy} + \Delta_x \Delta_y k~,
\fe
where $k$ is an integer-valued gauge parameter on the sites.

We suppress the ``vortices'' by modifying the Villain action \eqref{XYplaq-Vill-action} as
\ie\label{XYplaq-modVill-action}
\frac{\beta_0}{2} \sum_{\tau\text{-link}} (\Delta_\tau \phi - 2\pi n_\tau)^2 + \frac{\beta}{2} \sum_{xy\text{-plaq}} (\Delta_x \Delta_y \phi - 2\pi n_{xy})^2 + i \sum_\text{cube} \phi^{xy} (\Delta_\tau n_{xy} - \Delta_x \Delta_y n_\tau)~,
\fe
 where $\phi^{xy}$ is a real Lagrange multiplier field on the cubes or dual sites of the lattice.  It imposes $\Delta_\tau n_{xy} - \Delta_x \Delta_y n_\tau=0$, which can be interpreted as vanishing field strength of the gauge field $(n_\tau,n_{xy})$. We will refer to this and similar constraints as flatness constraints.  $\phi^{xy}$ has a gauge symmetry
\ie
\phi^{xy} \sim \phi^{xy} + 2\pi k^{xy}~,
\fe
where $k^{xy}$ is an integer-valued gauge parameter on the cubes of the lattice.
We will refer to \eqref{XYplaq-modVill-action} as the modified Villain version of \eqref{XYplaq-action}.

\subsubsection{Self-Duality}
Using the Poisson resummation formula \eqref{Possonresummationi}, we can dualize the modified Villain action \eqref{XYplaq-modVill-action} to
\ie\label{XYplaq-modVill-dualaction}
& \frac{1}{2(2\pi)^2\beta} \sum_{\text{dual }\tau\text{-link}} (\Delta_\tau \phi^{xy} - 2\pi n^{xy}_\tau)^2 + \frac{1}{2(2\pi)^2\beta_0} \sum_{\text{dual }xy\text{-plaq}} (\Delta_x \Delta_y \phi^{xy} - 2\pi n)^2
\\
&\qquad \qquad\qquad\qquad - i \sum_\text{site} \phi (\Delta_\tau n - \Delta_x \Delta_y n^{xy}_\tau)~,
\fe
where $n^{xy}_\tau$ and $n$ are integer-valued fields on the dual $\tau$-links and the dual $xy$-plaquettes respectively. We interpret $(n^{xy}_\tau,n)$ as $\mathbb Z$ tensor gauge fields that make $\phi^{xy}$ compact because of the gauge symmetry
\ie\label{XYplaq-modVill-windgaugesym}
&\phi^{xy} \sim \phi^{xy} + 2\pi k^{xy}~,
\\
&n^{xy}_\tau \sim n^{xy}_\tau + \Delta_\tau k^{xy}~,
\\
&n \sim n + \Delta_x \Delta_y k^{xy}~.
\fe
Here, the field $\phi$ is a Lagrange multiplier that imposes the constraint that the gauge invariant field strength of $(n^{xy}_\tau,n)$ vanishes; i.e., it is flat.
Therefore, the modified Villain model \eqref{XYplaq-modVill-action} is self-dual with $\beta_0 \leftrightarrow \frac{1}{(2\pi)^2\beta}$.

\subsubsection{Global symmetries}
In all the three models, \eqref{XYplaq-action}, \eqref{XYplaq-Vill-action}, and \eqref{XYplaq-modVill-action}, there is a \emph{$(\mathbf 1_0,\mathbf 1_2)$ momentum dipole symmetry}, which acts on the fields as
\ie
\phi \rightarrow \phi + c^x(\hat x) + c^y(\hat y)~,
\fe
where $c^i(\hat x^i)$ is real-valued. Due to the zero mode of the gauge symmetry \eqref{XYplaq-Vill-gaugesym}, the momentum dipole symmetry is $U(1)$. Using \eqref{XYplaq-modVill-action}, the components of the Noether current of the momentum dipole symmetry are
\ie
&J_\tau = i\beta_0 (\Delta_\tau \phi - 2\pi n_\tau) = \frac{1}{2\pi}(\Delta_x \Delta_y \phi^{xy} - 2\pi n)~,
\\
&J^{xy} = i\beta (\Delta_x \Delta_y \phi - 2\pi n_{xy}) = \frac{1}{2\pi}(\Delta_\tau \phi^{xy} - 2\pi n^{xy}_\tau)~.
\fe
$(J_\tau,J^{xy})$ are in the $(\mathbf 1_0,\mathbf 1_2)$ representations of $\mathbb{Z}_4$. They satisfy the $(\mathbf 1_0,\mathbf 1_2)$ dipole conservation equation
\ie
\Delta_\tau J_\tau = \Delta_x \Delta_y J^{xy}~,
\fe
because of the equation of motion of $\phi$. The momentum dipole charges are
\ie
Q^x(\hat x, \tilde{\mathcal C}^x) &= \sum_{\text{dual }xy\text{-plaq}\in \tilde{\mathcal C}^x} J_\tau + \sum_{\text{dual }\tau x\text{-plaq}\in \tilde{\mathcal C}^x} \Delta_x J^{xy}~,
\\
&=-\sum_{\text{dual }xy\text{-plaq}\in \tilde{\mathcal C}^x} n - \sum_{\text{dual }\tau x\text{-plaq}\in \tilde{\mathcal C}^x} \Delta_x n^{xy}_\tau
\fe
where $\tilde{\mathcal C}^x$ is a strip along the dual $xy$- and $\tau x$-plaquettes in the $\tau y$ plane at fixed $\hat x$. The second line can be interpreted as the Wilson ``strip'' operator of $(n^{xy}_\tau,n)$. Similarly, we can define $Q^y(\hat y, \tilde{\mathcal C}^y)$. When $\tilde{\mathcal C}^x$ and $\tilde{\mathcal C}^y$ are purely spatial at a fixed $\hat \tau$, the charges satisfy the constraint
\ie
\sum_{\hat x\text{: fixed }\hat \tau} Q^x(\hat x) = \sum_{\hat y\text{: fixed }\hat \tau} Q^y(\hat y) = \sum_{\text{dual }xy\text{-plaq: fixed }\hat \tau} J_\tau~.
\fe
The charged momentum operators are $e^{i\phi}$.

The modified Villain model \eqref{XYplaq-modVill-action} also has a \emph{$(\mathbf 1_2,\mathbf 1_0)$ winding dipole symmetry}, which acts on the fields as
\ie
\phi^{xy} \rightarrow \phi^{xy} + c^{xy}_x(\hat x) + c^{xy}_y(\hat y)~,
\fe
where $c^{xy}_i(\hat x^i)$ is real-valued.
By contrast, this symmetry is absent in the original lattice model  \eqref{XYplaq-action} and its Villain version \eqref{XYplaq-Vill-action}.
Due to the zero mode of the gauge symmetry \eqref{XYplaq-modVill-windgaugesym}, the winding dipole symmetry is $U(1)$. The components of the Noether current of the winding dipole symmetry are
\ie
&J^{xy}_\tau = -\frac{i}{(2\pi)^2\beta} (\Delta_\tau \phi^{xy} - 2\pi n^{xy}_\tau) = \frac{1}{2\pi}(\Delta_x \Delta_y \phi - 2\pi n_{xy})~,
\\
&J = -\frac{i}{(2\pi)^2\beta_0} (\Delta_x \Delta_y \phi^{xy} - 2\pi n) = \frac{1}{2\pi}(\Delta_\tau \phi - 2\pi n_\tau)~.
\fe
They satisfy the $(\mathbf 1_2,\mathbf 1_0)$ dipole conservation equation
\ie
\Delta_\tau J^{xy}_\tau = \Delta_x \Delta_y J~,
\fe
because of the equation of motion of $\phi^{xy}$. The winding dipole charges are
\ie
Q^{xy}_x(\hat x, \mathcal C^x) &= \sum_{xy\text{-plaq}\in \mathcal C^x} J^{xy}_\tau + \sum_{\tau x\text{-plaq}\in \mathcal C^x} \Delta_x J~,
\\
&= -\sum_{xy\text{-plaq}\in \mathcal C^x} n_{xy} - \sum_{\tau x\text{-plaq}\in \mathcal C^x} \Delta_x n_\tau~,
\fe
where $\mathcal C^x$ is a strip along the $xy$- and $\tau x$-plaquettes in the $\tau y$ plane at fixed $\hat x$. The second line can be interpreted as the Wilson ``strip'' operator of $(n_\tau,n_{xy})$. Similarly, we can define $Q^{xy}_y(\hat y,\mathcal C^y)$. When $\mathcal C^x$ and $\mathcal C^y$ are purely spatial at a fixed $\hat \tau$, the charges satisfy the constraint
\ie
\sum_{\hat x\text{: fixed }\hat \tau} Q^{xy}_x(\hat x) = \sum_{\hat y\text{: fixed }\hat \tau} Q^{xy}_y(\hat y) = \sum_{xy\text{-plaq: fixed }\hat \tau} J^{xy}_\tau~.
\fe
The charged winding operators are $e^{i\phi^{xy}}$.

There is a mixed 't Hooft anomaly between the two $U(1)$ global symmetries. One way to see this is to couple the system to the classical background gauge fields $(A_\tau,A_{xy};N_{\tau xy})$ and $(\tilde A^{xy}_\tau,\tilde A;\tilde N_\tau)$ of the momentum and winding symmetries, respectively.
Here $A_\tau,A_{xy}, \tilde A^{xy}_\tau,\tilde A$ are real-valued and $N_{\tau xy} ,\tilde N_\tau$ are integer-valued.
(See a similar discussion in Appendix \ref{sec:XY-sym}.)
The action is:
\ie
&\frac{\beta_0}{2} \sum_{\tau\text{-link}} (\Delta_\tau \phi - A_\tau - 2\pi n_\tau)^2 + \frac{\beta}{2} \sum_{xy\text{-plaq}} (\Delta_x \Delta_y \phi - A_{xy} - 2\pi n_{xy})^2 \\
&\qquad+ i \sum_\text{cube} \phi^{xy} (\Delta_\tau n_{xy} - \Delta_x \Delta_y n_\tau + N_{\tau xy})
\\
& \qquad- \frac{i}{2\pi}\sum_{xy\text{-plaq}} \tilde A^{xy}_\tau (\Delta_x \Delta_y \phi - A_{xy} - 2\pi n_{xy}) - \frac{i}{2\pi}\sum_{\tau\text{-link}} \tilde A (\Delta_\tau \phi - A_\tau - 2\pi n_\tau) - i \sum_\text{site} \tilde N_\tau \phi ~,
\fe
with the gauge symmetry
\ie
&\phi \sim \phi + \alpha + 2\pi k~, && \phi^{xy} \sim \phi^{xy} + \tilde \alpha^{xy} + 2\pi k^{xy}~,
\\
&A_\tau \sim A_\tau + \Delta_\tau \alpha + 2\pi K_\tau~, && \tilde A^{xy}_\tau \sim \tilde A^{xy}_\tau + \Delta_\tau \tilde \alpha^{xy} + 2\pi \tilde K^{xy}_\tau~,
\\
&A_{xy} \sim A_{xy} + \Delta_x \Delta_y \alpha + 2\pi K_{xy}~, && \tilde A \sim \tilde A + \Delta_x \Delta_y \tilde \alpha^{xy} + 2\pi \tilde K~,
\\
&n_\tau \sim n_\tau + \Delta_\tau k - K_\tau~, && \tilde N_\tau \sim \tilde N_\tau + \Delta_\tau \tilde K - \Delta_x \Delta_y \tilde K^{xy}_\tau~.
\\
&n_{xy} \sim n_{xy} + \Delta_x \Delta_y k - K_{xy}~,
\\
&N_{\tau xy} \sim N_{\tau xy} + \Delta_\tau K_{xy} - \Delta_x \Delta_y K_\tau~,\qquad
\fe
Here, $K_\tau, K_{xy}, \tilde K^{xy}_\tau,\tilde K$ are integers, and $\alpha,\tilde \alpha^{xy}$ are real.  They are the classical gauge parameters of the classical background gauge fields $(A_\tau,A_{xy};N_{\tau xy})$ and $(\tilde A^{xy}_\tau,\tilde A;\tilde N_\tau)$ . The variation of the action under the gauge transformation is
\ie\label{XYplaq-anomaly-lattice}
&- \frac{i}{2\pi}\sum_\text{site} \tilde \alpha^{xy} (\Delta_\tau A_{xy} - \Delta_x \Delta_y A_\tau - 2\pi N_{\tau xy})
\\
&\quad+ i\sum_{xy\text{-plaq}} \tilde K^{xy}_\tau (A_{xy}+\Delta_x\Delta_y\alpha) + i\sum_{\tau\text{-link}} \tilde K (A_\tau+\Delta_\tau\alpha)-i\sum_{\text{site}} \tilde N_\tau \alpha~.
\fe
It signals an anomaly because it cannot be cancelled by adding to the action any 2+1d local counterterms.

\subsubsection{A convenient gauge choice}\label{sec:2+1phicontinuum}

We now discuss a convenient gauge choice that sets most of the integer gauge fields to zero.
We first integrate out $\phi^{xy}$, which imposes the flatness condition on $(n_\tau,n_{xy})$.
We then gauge fix $n_\tau=0$ and $n_{xy}=0$ except for $n_\tau(L^\tau-1,\hat x, \hat y)$, $n_{xy}(\hat \tau, \hat x, L^y-1)$, and $n_{xy}(\hat \tau,L^x-1,\hat y)$. The remaining gauge-invariant information is in the holonomies:
\ie\label{XYplaq-gaugefixing}
&n_\tau(L^\tau-1,\hat x, \hat y) = \bar n^x(\hat x) + \bar n^y(\hat y)~,
\\
&n_{xy}(\hat \tau, \hat x, L^y-1) = \bar n^{xy}_x(\hat x)~,
\\
&n_{xy}(\hat \tau,L^x-1,\hat y) = \bar n^{xy}_y(\hat y)~,
\fe
where $\bar n^i(\hat x^i)$ and $\bar n^{xy}_i(\hat x^i)$ are integer-valued. There is a gauge ambiguity in the zero modes of $\bar n^i(\hat x^i)$, while $\bar n^{xy}_i(\hat x^i)$ satisfy the constraint $\bar n^{xy}_x(L^x-1) = \bar n^{xy}_y(L^y-1)$. In total, there are $2L^x+2L^y-2$ independent integers that cannot be gauged away. The residual gauge symmetry is
\ie\label{residgaphi}
\phi \sim \phi + 2\pi w^x(\hat x) + 2\pi w^y(\hat y)~,
\fe
where $w^i(\hat x^i)$ is integer-valued.

Let us define a new field $\bar \phi$ on the sites such that in the fundamental domain
\ie
\bar \phi(\hat \tau,\hat x,\hat y) = \phi(\hat \tau,\hat x,\hat y)~,\qquad \text{for }0\le \hat x^\mu < L^\mu~,
\fe
and beyond the fundamental domain, it is extended via
\ie\label{XYplaq-phibar}
&\bar \phi(\hat \tau+L^\tau,\hat x,\hat y) = \bar \phi(\hat \tau,\hat x,\hat y) - 2\pi \bar n^x(\hat x) - 2\pi \bar n^y(\hat y)~,
\\
&\bar \phi(\hat \tau,\hat x+L^x,\hat y) = \bar \phi(\hat \tau,\hat x,\hat y) - 2\pi \sum_{\hat y'=0}^{\hat y-1}\bar n^{xy}_y(\hat y')~,
\\
&\bar \phi(\hat \tau,\hat x,\hat y+L^y) = \bar \phi(\hat \tau,\hat x,\hat y) - 2\pi \sum_{\hat x'=0}^{\hat x-1}\bar n^{xy}_x(\hat x')~.
\fe
In particular, in the gauge \eqref{XYplaq-gaugefixing}, $\Delta_\tau \bar \phi = \Delta_\tau \phi - 2\pi n_\tau$, and  $\Delta_x \Delta_y \bar \phi = \Delta_x \Delta_y \phi - 2\pi n_{xy}$. Although $\phi$ and $(n_\tau,n_{xy})$ are single-valued, $\bar \phi$ can wind around the nontrivial cycles of spacetime. So, in the path integral, we should sum over nontrivial winding sectors of $\bar \phi$.
The action \eqref{XYplaq-modVill-action} in terms of $\bar \phi$ is
\ie\label{XYplaq-modVill-gaugefix-action}
\frac{\beta_0}{2} \sum_{\tau\text{-link}} (\Delta_\tau \bar \phi)^2 + \frac{\beta}{2} \sum_{xy\text{-plaq}} (\Delta_x \Delta_y \bar \phi)^2~.
\fe

Let us discuss some charged configurations in the lattice model \eqref{XYplaq-modVill-gaugefix-action}.  We define the periodic Kronecker delta function
\ie
\delta^P(\hat x,\hat x_0,L^x) \equiv \sum_{I \in \mathbb Z} \delta_{\hat x, \hat x_0 - I L^x}~,
\fe
and a suitable step function $\Theta^P(\hat x,\hat x_0,L^x)$ such that
\ie
\Theta^P(0,\hat x_0,L^x) = 0~,\qquad \Delta_x \Theta^P(\hat x,\hat x_0,L^x) = \delta^P(\hat x, \hat x_0, L^x)~.
\fe
Note that this function is not periodic in $\hat x$.  A minimal winding configuration is
\ie\label{XYplaq-modVill-minwind}
\bar \phi(\hat \tau,\hat x,\hat y) = 2\pi \left[ \frac{\hat x}{L^x} \Theta^P(\hat y,\hat y_0,L^y) + \frac{\hat y}{L^y} \Theta^P(\hat x,\hat x_0,L^x) - \frac{\hat x \hat y}{L^x L^y} \right]~.
\fe
The most general winding configuration can be obtained by taking linear combinations with integer coefficients of \eqref{XYplaq-modVill-minwind} with different $\hat x_0,\hat y_0$ and adding to it a periodic function.  The winding charges of \eqref{XYplaq-modVill-minwind} are $Q^{xy}_x(\hat x) = \delta^P(\hat x,\hat x_0,L^x)$ and $Q^{xy}_y(\hat y) = \delta^P(\hat y,\hat y_0,L^y)$.
This configuration satisfies the equation of motion of $\bar \phi$, so it is a minimal action configuration with these winding charges. Its action is
\ie
 \frac{\beta (2\pi)^2}{2} L^\tau \left( \frac{1}{L^x} + \frac{1}{L^y} - \frac{1}{L^xL^y} \right)
~.
\fe
Its Lorentzian interpretation is a winding state with energy
\ie
\frac{\beta (2\pi)^2}{2a} \left( \frac{1}{L^x} + \frac{1}{L^y} - \frac{1}{L^xL^y} \right)
~,
\fe
where $a$ is the lattice spacing.

\subsubsection{Continuum limit}

In the continuum limit, we take $a\rightarrow 0$, $L^\mu \to \infty $ with fixed $\ell^\mu = aL^\mu$.  In order for the limit to be nontrivial, we take the coupling constants to scale as $\beta_0 = \mu_0 a$ and $\beta = \frac{1}{\mu a}$.  Then, the action becomes
\ie\label{XYplaq-continuum-action}
\int d\tau dx dy~ \left[ \frac{\mu_0}{2} (\partial_\tau \phi)^2 + \frac{1}{2\mu} (\partial_x \partial_y \phi)^2 \right]~,
\fe
where we dropped the bar on $\phi$.
This is the Euclidean version of the 2+1d $\phi$-theory of \cite{paper1}, which had been first introduced in \cite{PhysRevB.66.054526}.
(See also \cite{Tay_2011,You:2019cvs,You:2019bvu,Karch:2020yuy,You:2021tmm}  for related discussions on this theory.)

The mixed 't Hooft anomaly between the momentum and winding symmetries can be seen by coupling the system to their background gauge fields $(A_\tau,A_{xy})$ and $(\tilde A^{xy}_\tau,\tilde A)$ respectively:
\ie
\int d\tau dx dy~ \left[ \frac{\mu_0}{2} (\partial_\tau \phi - A_\tau)^2 + \frac{1}{2\mu} (\partial_x \partial_y \phi - A_{xy})^2 - \frac{i}{2\pi} \tilde A^{xy}_\tau (\partial_x \partial_y \phi - A_{xy}) - \frac{i}{2\pi} \tilde A(\partial_\tau \phi - A_\tau) \right]~,
\fe
with gauge symmetry
\ie
&\phi \sim \phi + \alpha~, && \phi^{xy} \sim \phi^{xy} + \tilde \alpha^{xy}~,
\\
&A_\tau \sim A_\tau + \partial_\tau \alpha~, && \tilde A^{xy}_\tau \sim \tilde A^{xy}_\tau + \partial_\tau \tilde \alpha^{xy}~,
\\
&A_{xy} \sim A_{xy} + \partial_x \partial_y \alpha~, \qquad && \tilde A \sim \tilde A + \partial_x \partial_y \tilde \alpha^{xy}~.
\fe
Here, $\alpha, \tilde \alpha^{xy}$ are the gauge parameters. The variation of the action under the gauge transformation is
\ie
-\frac{i}{2\pi} \int d\tau dx dy~ \tilde \alpha^{xy} (\partial_\tau A_{xy} - \partial_x \partial_y A_\tau)~.
\fe
It signals an anomaly because it cannot be cancelled by adding to the action any 2+1d local counterterms. This is the continuum counterpart of the corresponding lattice expression \eqref{XYplaq-anomaly-lattice}.

We can also view the modified Villain lattice model \eqref{XYplaq-modVill-action}, or its gauge fixed version \eqref{XYplaq-modVill-gaugefix-action}, as a discretized version the continuum theory \eqref{XYplaq-continuum-action}.  Our analysis of this lattice model makes rigorous the various assertions in \cite{paper1}.  Let us discuss them in more detail.

Both the continuum theory \eqref{XYplaq-continuum-action} and the lattice theory \eqref{XYplaq-modVill-gaugefix-action} have real-valued fields and the periodicity in field space is implemented using the twisted boundary conditions \eqref{XYplaq-phibar}.

One could question whether the lattice theory \eqref{XYplaq-modVill-gaugefix-action} with this particular sum over twisted boundary conditions is fully consistent.  In the continuum, this was discussed in detail in \cite{paper1, Rudelius:2020kta}. On the lattice, the consistency follows from relating it to the lattice gauge theory \eqref{XYplaq-modVill-action} before the gauge fixing \eqref{XYplaq-gaugefixing}.  Furthermore, the remaining gauge freedom \eqref{residgaphi} in the lattice theory \eqref{XYplaq-modVill-gaugefix-action} can now be interpreted as the gauge freedom of the continuum theory \cite{paper1, Rudelius:2020kta}.

The discussion of \cite{paper1} uncovered a number of surprising properties of the continuum theory \eqref{XYplaq-continuum-action}, which are not present in the original microscopic theory \eqref{XYplaq-action}.  It has an emergent global dipole $U(1)$ winding symmetry and it is self dual.  Now we see these properties already in the modified Villain  lattice model \eqref{XYplaq-modVill-action}.  A reader who was skeptical about the continuum analysis of \cite{paper1} can be reassured by seeing it derived on the lattice.

For fixed $\ell^\tau$ and $\ell \sim \ell^x,\ell^y$, the action of the winding configuration \eqref{XYplaq-modVill-minwind} scales as $\ell^\tau/\mu \ell a$, which diverges as $1/ a$ in the continuum limit. The configuration \eqref{XYplaq-modVill-minwind} gives a precise meaning to the winding configuration with infinite action in the continuum \cite{paper1}.\footnote{The discussion of such infinite action and infinite energy configurations was described in \cite{paper1} as ``ambitious.''  It is rigorous in the context of the modified Villain model.}  More generally, the classification of discontinuous configurations in the continuum theory \eqref{XYplaq-continuum-action} \cite{paper1} is exactly as in the previous subsection.

In conclusion, the lattice model \eqref{XYplaq-modVill-action} flows in the continuum limit to \eqref{XYplaq-continuum-action}.  Conversely, the lattice model \eqref{XYplaq-modVill-action}, or its gauge fixed version \eqref{XYplaq-modVill-gaugefix-action},  gives a rigorous setting for the discussion of the continuum theory \eqref{XYplaq-continuum-action} of \cite{paper1}.

\subsection{$A$-theory ($U(1)$ tensor gauge theory)}
We can gauge the $U(1)$ momentum dipole symmetry by coupling \eqref{XYplaq-modVill-action} to the $(\mathbf 1_0,\mathbf 1_2)$ tensor gauge fields $(A_\tau,A_{xy})$. We will consider this system in Section \ref{sec:2+1d-ZN}, and restrict to the pure tensor gauge theory in this section.  This pure gauge theory was discussed on the lattice and in the continuum in \cite{paper1} (see also earlier work in \cite{Bulmash:2018lid,You:2019cvs,You:2019bvu,Dubinkin:2020kxo}.

We place the $U(1)$ variables $e^{iA_\tau}$ and $e^{iA_{xy}}$ on $\tau$-links and $xy$-plaquettes of the lattice respectively. The action for the pure $U(1)$ tensor gauge theory is
\ie\label{2+1d-U1-tensor-action}
\gamma \sum_\text{cube} [1-\cos(\Delta_\tau A_{xy} - \Delta_x \Delta_y A_\tau)]~,
\fe
where $A_\tau$ and $A_{xy}$ are circle-valued fields.  It has the gauge symmetry
\ie\label{2+1d-U1-tensor-gaugesym}
&e^{iA_\tau} \sim e^{iA_\tau + i\Delta_\tau \alpha}~,
\\
&e^{iA_{xy} }\sim e^{iA_{xy} + i\Delta_x \Delta_y \alpha} ~,
\fe
with circle-valued $\alpha$ on the sites.

At large $\gamma$, we can approximate \eqref{2+1d-U1-tensor-action} by the Villain action
\ie\label{2+1d-U1-tensor-Vill-action}
\frac{\gamma}{2} \sum_\text{cube} (\Delta_\tau A_{xy} - \Delta_x \Delta_y A_\tau - 2\pi n_{\tau xy})^2~,
\fe
where $n_{\tau xy}$ is an integer-valued field on the cubes.  Now we view the gauge fields $(A_\tau,A_{xy})$ and the gauge parameters $\alpha$ as real-valued, and the gauge symmetry \eqref{2+1d-U1-tensor-gaugesym} becomes
\ie\label{2+1d-U1-tensor-Vill-gaugesym}
&A_\tau \sim A_\tau + \Delta_\tau \alpha + 2\pi k_\tau~,
\\
&A_{xy} \sim A_{xy} + \Delta_x \Delta_y \alpha + 2\pi k_{xy}~,
\\
&n_{\tau xy} \sim n_{\tau xy} + \Delta_\tau k_{xy} - \Delta_x \Delta_y k_\tau~,
\fe
where the gauge parameters $k_\tau$ and $k_{xy}$ are integers on the $\tau$-links and $xy$-plaquettes respectively.

We can interpret $n_{\tau xy}$ as the $\mathbb Z$ gauge field that makes $(A_\tau,A_{xy})$ compact. In contrast to the XY-plaquette model, the $U(1)$ tensor gauge theory has no ``vortices.'' So, we do not modify the Villain action \eqref{2+1d-U1-tensor-Vill-action} as in \eqref{XYplaq-modVill-action}.  Indeed, there is no local gauge-invariant field strength constructed out of the gauge field $n_{\tau xy}$.

We can also add a $\theta$-term to the Villain action \eqref{2+1d-U1-tensor-Vill-action}:
\ie\label{2+1d-U1-tensor-Vill-theta-action}
\frac{\gamma}{2} \sum_\text{cube} E_{xy}^2 + \frac{i\theta}{2\pi} \sum_\text{cube}E_{xy}~,
\fe
where we defined the electric field
\ie
E_{xy} = \Delta_\tau A_{xy} - \Delta_x \Delta_y A_\tau - 2\pi n_{\tau xy}~,
\fe
on the cubes. Since $(A_\tau,A_{xy})$ is single-valued, we can write the $\theta$-term as $-i\theta \sum_\text{cube} n_{\tau xy}$, which implies that the theta angle is $2\pi$-periodic, i.e., $\theta \sim \theta+2\pi$.  Note that such a $\theta$-term cannot be added in the original formulation \eqref{2+1d-U1-tensor-action}, while it is straightforward and natural in the Villain version \eqref{2+1d-U1-tensor-Vill-action}.

The quantized electric fluxes
\ie
&e^x(\hat x)=\sum_{\text{cube: fixed }\hat x} E_{xy} = -2\pi \sum_{\text{cube: fixed }\hat x} n_{\tau xy}\in 2\pi \mathbb Z~,
\\
&e^y(\hat y)=\sum_{\text{cube: fixed }\hat y} E_{xy} = -2\pi \sum_{\text{cube: fixed }\hat y} n_{\tau xy}\in 2\pi \mathbb Z~,
\fe
are associated with nontrivial holonomies of $n_{\tau xy}$ and they characterize the bundles of the tensor gauge theory. These fluxes satisfy the constraint
\ie
\sum_{\hat x} e^x(\hat x) = \sum_{\hat y} e^y(\hat y) = \sum_\text{cube} E_{xy}~.
\fe

\subsubsection{Global symmetries}
The three models \eqref{2+1d-U1-tensor-action}, \eqref{2+1d-U1-tensor-Vill-action}, and \eqref{2+1d-U1-tensor-Vill-theta-action} have an \emph{electric tensor symmetry} that acts on the fields as
\ie
A_\tau \rightarrow A_\tau + \lambda_\tau~,\qquad A_{xy} \rightarrow A_{xy} + \lambda_{xy}~,
\fe
where $(\lambda_\tau,\lambda_{xy})$ is a flat, real-valued tensor gauge field (i.e., it has vanishing field strength).\footnote{Using the $\alpha$ gauge symmetry of \eqref{2+1d-U1-tensor-Vill-gaugesym}, and the flatness of $(\lambda_\tau,\lambda_{xy})$, we can set $\lambda_\tau=c^x(\hat x)+c^y(\hat y)$, and  $\lambda_{xy} = c^{xy}_x(\hat x) + c^{xy}_y(\hat y)$, where $c^i(\hat x^i)$ and $c^{xy}_i(\hat x^i)$ are real-valued.} Due to the integer-valued gauge symmetry with $(k_\tau,k_{xy})$ \eqref{2+1d-U1-tensor-Vill-gaugesym}, the electric tensor symmetry is $U(1)$, rather than $\bR$. The Noether current of this electric symmetry follows from \eqref{2+1d-U1-tensor-Vill-theta-action}
\ie
J^{xy}_\tau = -i \gamma E_{xy} + \frac{\theta}{2\pi}~.
\fe
It satisfies the conservation equation and the differential condition (Gauss law)
\ie
\Delta_\tau J^{xy}_\tau = 0~,\qquad \Delta_x \Delta_y J^{xy}_\tau = 0~,
\fe
due to the equations of motion of $A_{xy}$ and $A_\tau$ respectively. The conserved charge is
\ie
Q(\hat x, \hat y) = J^{xy}_\tau = Q^x(\hat x) + Q^y(\hat y)~,
\fe
where $Q^i(\hat x^i)$ is an integer, and the second equation follows from the Gauss law. The observables charged under the electric symmetry are the Wilson defect/operator
\ie
&W^\tau(\hat x,\hat y) = \exp\left[ i \sum_{\tau\text{-link: fixed }\hat x,\hat y} A_\tau \right]~,
\\
&W^x(\hat x,\mathcal C^x) = \exp\left[ i \sum_{xy\text{-plaq}\in\mathcal C^x} A_{xy} + i \sum_{\tau x\text{-plaq}\in\mathcal C^x} \Delta_x A_\tau \right]~,
\fe
where $\mathcal C^x$ is a closed strip along the $xy$- and $\tau x$-plaquettes in the $\tau y$-plane at a fixed $\hat x$. Similarly, there is $W^y(\hat y,\mathcal C^y)$.

\subsubsection{Gauge-fixing and the continuum limit}\label{sec:2+1Acont}

Using the integer gauge freedom \eqref{2+1d-U1-tensor-Vill-gaugesym}, we gauge fix $n_{\tau xy}=0$, except for
\ie
n_{\tau xy}(L^\tau-1,\hat x,L^y-1) \equiv \bar n^x_{\tau xy}(\hat x)~,\qquad n_{\tau xy}(L^\tau -1,L^x-1,\hat y)\equiv \bar n^y_{\tau xy}(\hat y)~.
\fe
The integers $\bar n^i_{\tau xy}(\hat x^i)$ capture the only gauge-invariant information in $n_{\tau xy}$: its holonomies. They satisfy a constraint $\bar n^x_{\tau xy}(L^x-1)=\bar n^y_{\tau xy}(L^y-1)$. The residual gauge freedom is
\ie
&A_\tau \sim A_\tau + \Delta_\tau \alpha + 2\pi k_\tau~,
\\
&A_{xy} \sim A_{xy} + \Delta_x \Delta_y \alpha + 2\pi k_{xy}~,
\fe
where $(k_\tau,k_{xy})$ is a flat, integer-valued tensor gauge field.

Similar to \eqref{XYplaq-phibar} in the $\phi$-theory, we  define a new tensor gauge field $(\bar A_\tau,\bar A_{xy})$ on the $\tau$-links and $xy$-plaquettes such that
\ie
\Delta_\tau \bar A_{xy} - \Delta_x \Delta_y \bar A_\tau = \Delta_\tau A_{xy} - \Delta_x \Delta_y A_\tau - 2\pi n_{\tau xy}~.
\fe
Although $(A_\tau,A_{xy})$ and $n_{\tau xy}$ are single-valued, $(\bar A_\tau,\bar A_{xy})$ can have nontrivial monodromies  around nontrivial cycles of the Euclidean spacetime. So, in the path integral, we should sum over nontrivial twisted sectors of $(\bar A_\tau,\bar A_{xy})$.

The action \eqref{2+1d-U1-tensor-Vill-theta-action} in terms of $(\bar A_\tau,\bar A_{xy})$ is
\ie\label{2+1d-U1-tensor-Vill-theta-action-gaugefixed}
\frac{\gamma}{2} \sum_\text{cube} \bar E_{xy}^2 + \frac{i\theta}{2\pi} \sum_\text{cube} \bar E_{xy}~,
\fe
where we defined the new electric field
\ie
\bar E_{xy} = \Delta_\tau \bar A_{xy} - \Delta_x \Delta_y \bar A_\tau~,
\fe
on the cubes.

In the continuum limit $a\rightarrow 0$, choosing the coupling to scale as $\gamma = \frac{2}{a^3g_e^2}$ and the fields to scale as $\bar A_\tau = a A_\tau$ and $\bar A_{xy} = a^2 A_{xy}$,\footnote{The continuum tensor gauge fields $(A_\tau,A_{xy})$ and their electric field defined here are not the same as the ones defined on the lattice at the beginning of this section. We hope this does not cause any confusion.} the action becomes
\ie\label{2+1d-U1-tensor-action-contlimit}
&\int d\tau dx dy~\left( \frac{1}{g_e^2} E_{xy}^2 + \frac{i\theta}{2\pi} E_{xy} \right)~,\\
&E_{xy}=\partial_\tau A_{xy} - \partial_x \partial_y A_\tau~.
\fe
This is the Euclidean version of the continuum  2+1d $A$-theory of \cite{paper1}.
(See also \cite{Bulmash:2018lid,You:2019cvs,You:2019bvu,Dubinkin:2020kxo}.)
 The Villain model \eqref{2+1d-U1-tensor-Vill-theta-action} has the same $U(1)$ electric symmetry as the continuum $A$-theory.

The spectrum of the lattice model consists of light states, whose action scales as $a$. In the continuum limit $a\rightarrow 0$ with fixed $\ell^\tau$, $\ell^x$ and $\ell^y$, these light states become infinitely degenerate. The details can be found in \cite{paper1}.

We conclude that the lattice model \eqref{2+1d-U1-tensor-Vill-theta-action} flows in the continuum limit to \eqref{2+1d-U1-tensor-action-contlimit}.  Conversely, the lattice model \eqref{2+1d-U1-tensor-Vill-theta-action}, or its gauge fixed version \eqref{2+1d-U1-tensor-Vill-theta-action-gaugefixed},   give a rigorous setting for the discussion of the continuum theory \eqref{2+1d-U1-tensor-action-contlimit} of \cite{paper1}.

\subsection{$\mathbb{Z}_N$ tensor gauge theory}\label{sec:2+1d-ZN}

In this subsection,
we will consider the  modified Villain lattice version of the 2+1d $\bZ_N$ Ising plaquette model \cite{plaqising}.
The modified Villain lattice model takes the form of  a $BF$-type action, which admits two equivalent presentations.
The first one, which we call the integer $BF$-action, uses  only  integer-valued fields, while the second one, which we call the real $BF$-action, uses both real and integer-valued fields.
The real $BF$-action is  naturally connected to the continuum $\bZ_N$ tensor gauge theory of \cite{paper1}.

We can restrict the $U(1)$ variables in the $U(1)$ tensor gauge theory \eqref{2+1d-U1-tensor-action} to $\mathbb{Z}_N$ variables $e^{iA_\tau}=e^{\frac{2\pi i}{N} m_\tau}$ and $e^{iA_{xy}}=e^{\frac{2\pi i}{N} m_{xy}}$ with integers $m_\tau$ and $m_{xy}$. This leads to the $\mathbb{Z}_N$ tensor gauge theory with the action
\ie\label{2+1d-ZN-tensor-action}
\gamma \sum_\text{cube} \left[1-\cos\left(\frac{2\pi}{N}(\Delta_\tau m_{xy} - \Delta_x \Delta_y m_\tau)\right)\right]~.
\fe
At large $\gamma$, $\Delta_\tau m_{xy} - \Delta_x \Delta_y m_\tau=0$ mod $N$ and we can replace the action by
\ie\label{2+1d-ZN-tensor-Vill-action}
\frac{2\pi i}{N} \sum_\text{cube} \tilde m^{xy}(\Delta_\tau m_{xy} - \Delta_x \Delta_y m_\tau )~,
\fe
where $\tilde m^{xy}$ is an integer-valued field on the cubes.
We will refer to this presentation of the $\bZ_N$ tensor gauge theory as the \textit{integer $BF$-action}.
This is analogous to the presentation \eqref{eq:ZNtopoaction} for the topological lattice $\bZ_N$ gauge theory reviewed in Appendix \ref{pformgau}.

There is a gauge symmetry
\ie\label{2+1d-ZN-tensor-Vill-gaugesym}
&m_\tau \sim m_\tau + \Delta_\tau \ell + N k_\tau~,
\\
&m_{xy} \sim m_{xy} + \Delta_x \Delta_y \ell + N k_{xy}~,
\\
&\tilde m^{xy} \sim \tilde m^{xy} + N\tilde k^{xy}~,
\fe
where $\ell$ is an integer-valued field on the sites, $k_\tau$ and $k_{xy}$ are integer-valued fields on the $\tau$-links and $xy$-plaquettes respectively, and $\tilde k^{xy}$ is an integer-valued field on the cubes.

\subsubsection{Global symmetries}

In both models, \eqref{2+1d-ZN-tensor-action} and \eqref{2+1d-ZN-tensor-Vill-action}, there is a $\mathbb{Z}_N$ \emph{electric tensor symmetry}, which shifts $(m_\tau,m_{xy})$ by a flat, integer-valued tensor gauge field. In the presentation of the model based on \eqref{2+1d-ZN-tensor-Vill-action}, the charge operator is
\ie
U(\hat \tau,\hat x,\hat y)=\exp\left[\frac{2\pi i}{N}\tilde m^{xy}\right]~.
\fe
The observables charged under the electric symmetry are the Wilson defect/operator
\ie
&W^\tau(\hat x,\hat y) = \exp\left[ \frac{2\pi i}{N} \sum_{\tau\text{-link: fixed }\hat x,\hat y} m_\tau \right]~,
\\
&W^x(\hat x,\mathcal C^x) = \exp\left[ \frac{2\pi i}{N} \sum_{xy\text{-plaq}\in\mathcal C^x} m_{xy} + \frac{2\pi i}{N} \sum_{\tau x\text{-plaq}\in\mathcal C^x} \Delta_x m_\tau \right]~,
\fe
where $\mathcal C^x$ is a strip along the $xy$- and $\tau x$-plaquettes in the $\tau y$-plane at a fixed $\hat x$. Similarly, there is $W^y(\hat y,\mathcal C^y)$.

In the presentation of the model based on \eqref{2+1d-ZN-tensor-Vill-action}, but not in \eqref{2+1d-ZN-tensor-action}, there is also a $\mathbb{Z}_N$ magnetic dipole symmetry. The charge operators are $W^x(\hat x,\mathcal C^x)$ and $W^y(\hat y,\mathcal C^y)$, and the charged operator is $U(\hat \tau,\hat x,\hat y)$.

\subsubsection{Ground state degeneracy}

All the states of the model based on \eqref{2+1d-ZN-tensor-Vill-action} are degenerate.  The model has only ground states.  Let us count them. First, we sum over the integer-valued fields $m_\tau$ and $m_{xy}$. They impose the following constraint on $\tilde m^{xy}$
\ie
\Delta_\tau \tilde m^{xy}=\Delta_x\Delta_y \tilde m^{xy} =0\text{ mod } N~.
\fe
The gauge inequivalent configurations of $\tilde m^{xy}$ are
\ie
\tilde m^{xy}(\hat\tau,\hat x,\hat y)= \tilde m^{xy}_x(\hat x)+\tilde m^{xy}_y(\hat y)~,
\fe
where $\tilde m^{xy}_x(\hat x)$ and $\tilde m^{xy}_y(\hat y)$ are $\mathbb{Z}/N\mathbb{Z}$-valued. There is a gauge ambiguity in the zero modes of $\tilde m^{xy}_x(\hat x)$ and $\tilde m^{xy}_y(\hat y)$. So, in total, there are $N^{L^x+L^y-1}$ degenerate ground states.

\subsubsection{Real $BF$-action and the continuum limit}\label{sec:2+1realBF}

The model based on the integer $BF$-action \eqref{2+1d-ZN-tensor-Vill-action} has several different presentations.  Here we discuss a presentation in terms of real-valued and integer-valued fields, which is closer to the continuum limit.

We start with the integer $BF$-action \eqref{2+1d-ZN-tensor-Vill-action} and replace the integer-valued fields $\tilde m^{xy}$ and $(m_\tau,m_{xy})$ with real-valued fields $\tilde\phi^{xy}$ and $(A_\tau,A_{xy})$. In order to restrict these real-valued fields to be integer-valued, we add integer-valued Lagrange multiplier fields $n_{\tau xy}$ and $(\tilde n_\tau^{xy},\tilde n)$. Furthermore, since the gauge field $(A_\tau,A_{xy})$ has real-valued gauge symmetry, we introduce a real-valued Stueckelberg field $\phi$ for that gauge symmetry.
We end up with the action
\ie\label{2+1d-ZN-tensor-Vill-real-action}
&\frac{iN}{2\pi}\sum_{\text{cube}}\tilde \phi^{xy}(\Delta_\tau A_{xy}-\Delta_x\Delta_y A_\tau-2\pi n_{\tau xy})+iN\sum_{xy\text{-plaq}} A_{xy} \tilde n_\tau^{xy}
\\
&\quad +iN\sum_{\tau\text{-link}}A_\tau \tilde n+i\sum_{\text{site}}\phi(\Delta_\tau \tilde n-\Delta_x\Delta_y\tilde n_\tau^{xy})~,
\fe
where $\phi$, $\tilde \phi^{xy}$, $A_\tau$ and $A_{xy}$ are real-valued fields on the sites, the dual site, the $\tau$-links and the $xy$-plaquettes respectively, and $n_{\tau xy}$, $\tilde n_\tau^{xy}$ and $\tilde n$ are integer-valued fields on the cubes, the dual $\tau$-links, and the dual $xy$-plaquettes, respectively.

There action \eqref{2+1d-ZN-tensor-Vill-real-action} has the gauge symmetry
\ie\label{2+1d-ZN-tensor-Vill-real-gauge}
&\phi\sim\phi+N\alpha+2\pi k~,
\\
&\tilde \phi^{xy}\sim\tilde \phi^{xy}+2\pi \tilde k^{xy}~,
\\
&A_\tau\sim A_\tau +\Delta_\tau \alpha +2\pi k_\tau~,
\\
&A_{xy}\sim A_{xy}+\Delta_x\Delta_y\alpha+2\pi k_{xy}~,
\\
&\tilde n_\tau^{xy}\sim \tilde n_\tau^{xy}+\Delta_\tau \tilde k^{xy}~,
\\
&\tilde n\sim \tilde n+\Delta_x\Delta_y \tilde k^{xy}~,
\\
&n_{\tau xy}\sim n_{\tau xy}+\Delta_\tau k_{xy}-\Delta_x\Delta_y k_\tau~.
\fe

As a check, summing over the integer-valued fields $n_{\tau xy}$, $\tilde n_\tau^{xy}$, and $\tilde n$ in \eqref{2+1d-ZN-tensor-Vill-real-action} constrains
\ie
\tilde \phi^{xy}=\frac{2\pi}{N}\tilde m^{xy},\quad A_\tau - \frac{1}{N}\Delta_\tau \phi=\frac{2\pi}{N}m_\tau,\quad A_{xy} - \frac{1}{N}\Delta_x \Delta_y \phi=\frac{2\pi}{N}m_{xy}~,
\fe
where $\tilde m^{xy}$, $m_\tau$ and $m_{xy}$ are integer-valued fields. Substituting them back into the action leads to \eqref{2+1d-ZN-tensor-Vill-action}.

We  will refer to the  presentation \eqref{2+1d-ZN-tensor-Vill-real-action} of the $\bZ_N$ tensor gauge theory as the \textit{real $BF$-action}, which uses both  real and integer fields.
This is to be compared with the integer $BF$-action  \eqref{2+1d-ZN-tensor-action}, which uses only integer-valued fields.
These two presentations describe the same underlying lattice model, but use different sets of fields.
In the real $BF$-action, the integer fields effectively make the real fields compact.

The real $BF$-action \eqref{2+1d-ZN-tensor-Vill-real-action} can also be derived through Higgsing the $U(1)$ tensor gauge theory \eqref{2+1d-U1-tensor-Vill-theta-action} to a $\mathbb{Z}_N$ theory using the field $\phi$ in \eqref{XYplaq-modVill-action}. The Higgs action is
\ie
&\frac{i}{2\pi}\sum_{\tau\text{-link}} \tilde B(\Delta_\tau \phi-NA_\tau - 2\pi n_\tau) +  \frac{i}{2\pi}\sum_{xy\text{-plaq}} \tilde E^{xy}(\Delta_x \Delta_y \phi-NA_{xy} - 2\pi n_{xy})
\\
&\quad - i \sum_\text{cube} \tilde\phi^{xy} \left(\Delta_\tau n_{xy} - \Delta_x \Delta_y n_\tau+Nn_{\tau xy}\right)~,
\fe
where $\tilde B$ and $\tilde E^{xy}$ are real-valued fields on the $\tau$-links and the $xy$-plaquette, which implement the Higgsing as constraints. In addition to the gauge symmetry \eqref{2+1d-ZN-tensor-Vill-real-gauge}, there is a gauge symmetry
\ie
&n_\tau \sim n_\tau + \Delta_\tau k-Nk_\tau~,
\\
&n_{xy} \sim n_{xy} + \Delta_x \Delta_y k-Nk_{xy}~.
\fe
Summing over the integer-valued fields $n_\tau$ and $n_{xy}$ constrains
\ie
\tilde B-\Delta_x\Delta_y\tilde\phi^{xy}=-2\pi \tilde n~,\quad\tilde E^{xy}-\Delta_\tau\tilde \phi^{xy}=-2\pi\tilde n^{xy}_\tau~,
\fe
where $\tilde n$ and $\tilde n_\tau^{xy}$ are integer-valued fields. Substituting them back into the action leads to \eqref{2+1d-ZN-tensor-Vill-real-action}.

In a convenient gauge choice, most of the integer fields  are fixed to be zero, while the remaining ones enter into the twisted boundary conditions of the real fields.

Let us make it more explicit.
First, we integrate out $\phi$, which imposes the constraint $\Delta_\tau \tilde n-\Delta_x\Delta_y\tilde n_\tau^{xy}=0$. Then we can gauge fix $n_{\tau xy}$, $\tilde n_\tau^{xy}$ and $\tilde n$ to be zero almost everywhere except at
\ie\label{2+1d-ZN-tensor-gauge}
&n_{\tau xy}(L^\tau-1,\hat x,L^y-1)=\bar n_{\tau xy}^x(\hat x)~,
\\
&n_{\tau xy}(L^\tau-1,L^x-1,\hat y)=\bar n_{\tau xy}^y(\hat y)~,
\\
&\tilde n_{\tau}^{xy}(L^\tau-1,\hat x,\hat y)=\bar n_{\tau,x}^{xy}(\hat x)+\bar n_{\tau,y}^{xy}(\hat y)~,
\\
&\tilde n(\hat \tau,\hat x,L^y-1)=\bar n_{ x}(\hat x)~,
\\
&\tilde n(\hat \tau,L^x-1,\hat y)=\bar n_{ y}(\hat y)~,
\fe
where $\bar n_{\tau xy}^x,\bar n_{\tau xy}^y,\bar n_{\tau,x}^{xy},\bar n_{\tau,y}^{xy},\bar n_{ x},\bar n_{ y}$ are all integer-valued. These integers obey $\bar n_{\tau xy}^x(L^x-1)=\bar n_{\tau xy}^y(L^y-1)$ and $\bar n_{ x}(L^x-1)=\bar n_{ y}(L^y-1)$. The zero modes of $\bar n_{\tau,x}^{xy}(\hat x)$ and $\bar n_{\tau,y}^{xy}(\hat y)$ have a gauge ambiguity.

As in Sections \ref{sec:2+1phicontinuum} and \ref{sec:2+1Acont},
we define new fields $\bar \phi^{xy}$, $\bar A_\tau$ and $\bar A_{xy}$ on the sites, the $\tau$-links, and the $xy$-plaquettes such that
\ie
&\Delta_\tau \bar \phi^{xy}=\Delta_\tau \tilde \phi^{xy}-2\pi \tilde n_\tau^{xy}~,
\\
&\Delta_x\Delta_y \bar \phi^{xy}=\Delta_x\Delta_y \tilde \phi^{xy}-2\pi \tilde n~,
\\
&\Delta_\tau \bar A_{xy} - \Delta_x \Delta_y \bar A_\tau =\Delta_\tau A_{xy} - \Delta_x \Delta_y A_\tau - 2\pi n_{\tau xy}~.
\fe
In contrast to the original variables that are single-valued, the new variables can have nontrivial twisted boundary conditions around the nontrivial cycles of space-time.
So, in the path integral, we should sum over nontrivial  twisted  sectors of $\bar\phi^{xy}$ and $(\bar A_\tau,\bar A_{xy})$.

In terms of the new variables, the action \eqref{2+1d-ZN-tensor-Vill-real-action} becomes
\ie\label{2+1d-ZN-tensor-Vill-action-gaugefixed}
&\frac{iN}{2\pi}\sum_{\text{cube}}\bar \phi^{xy}(\Delta_\tau \bar A_{xy}-\Delta_x\Delta_y \bar A_\tau)+iN\sum_{\substack{xy\text{-plaq}\\\hat\tau = L^\tau-1}} \bar A_{xy} (\bar n_{\tau,x}^{xy}+\bar n_{\tau,y}^{xy})\\
&+iN\sum_{\substack{\tau\text{-link}\\ \hat x=L^x-1}}\bar A_\tau  \bar n_{ y}
+iN\sum_{\substack{\tau\text{-link}\\ \hat y=L^y-1}}\bar A_\tau  \bar n_{ x} - iN\sum_{\substack{\tau\text{-link}\\ \hat x=L^x-1\\\hat y=L^y-1}}\bar A_\tau  \bar n_{ x}~,
\fe

The real $BF$-action of our modified Villain model is closely related to the continuum field theory.
In the continuum limit, $a\rightarrow0$, the action becomes
\ie\label{2+1d-ZN-tensor-action-contlimit}
\frac{iN}{2\pi}\int d\tau dxdy~\phi^{xy}(\partial_\tau A_{xy}-\partial_x\partial_y A_\tau)~,
\fe
where we dropped the bars over the variables and rescaled them by appropriate powers of the lattice spacing $a$.  We also omitted the boundary terms that depend on the transition functions of $\phi^{xy}$ and $(A_\tau,A_{xy})$.\footnote{Such boundary terms are necessary in order to make the continuum action \eqref{2+1d-ZN-tensor-action-contlimit} well-defined.  They played a crucial role in the analysis of \cite{Rudelius:2020kta}.} This is the Euclidean version of the 2+1d $\mathbb{Z}_N$ tensor gauge theory of \cite{paper1}.

We conclude that the lattice model \eqref{2+1d-ZN-tensor-Vill-action}, or equivalently \eqref{2+1d-ZN-tensor-Vill-real-action}, flows in the continuum limit to \eqref{2+1d-ZN-tensor-action-contlimit}.  Conversely, the lattice model \eqref{2+1d-ZN-tensor-Vill-action},  or equivalently \eqref{2+1d-ZN-tensor-Vill-real-action},  gives a rigorous setting for the discussion of the continuum theory \eqref{2+1d-ZN-tensor-action-contlimit} of \cite{paper1}.

\section{3+1d (4d Euclidean) exotic theories with cubic symmetry}\label{sec:3+1d-cubic}
In this section, we will describe the modified Villain formulation of the exotic 3+1d continuum theories of \cite{paper2,paper3}. All the models are placed on a periodic 4d Euclidean lattice with lattice spacing $a$, and $L^\mu$ sites in the $\mu$ direction. We label the sites by integers $\hat x^\mu \sim \hat x^\mu + L^\mu$.

Since the spatial lattice has an $S_4$ rotation symmetry, we can organize the fields according to $S_4$ representations: the trivial representation $\mathbf{1}$, the sign representation $\mathbf{1}'$, a two-dimensional irreducible representation $\mathbf{2}$, the standard representation $\mathbf{3}$ and another three-dimensional irreducible representation $\mathbf{3}'$.

We will label the components of $S_4$ representations using $SO(3)$ vector indices $i,j,k$.
In this section, the indices $i,j, k$ in every expression are never equal, $i\neq j\neq k$.

We label the components of an object $V$ in $\mathbf 3$ of $S_4$ as $V_i$ and the components of an object $E$ in $\mathbf 3'$ of $S_4$ as $E_{ij}=E_{ji}$.  The labeling of the components of $T$ in $\mathbf 2$ of $S_4$ is slightly more complicated.
We can label them as $T^{[ij]k}=-T^{[ji]k}$, with an identification under simultaneous shifts of $T^{[xy]z}$, $T^{[yz]x}$, $T^{[zx]y}$ by the same amount. Alternatively, we can define the combinations $T^{k(ij)} = T^{[ki]j} - T^{[jk]i}$, which are not subject to the identification.  In this presentation, we have a constraint $T^{x(yz)}+T^{y(zx)}+T^{z(xy)}=0$. We will also use $T_{k(ij)}=T_{k(ji)}$ with lower indices to label the components of $\mathbf 2$. It has an identification under simultaneous shifts of $T_{x(yz)}$, $T_{y(zx)}$, $T_{z(xy)}$ by the same amount. Similarly, we define the combinations $T_{[ij]k} =  T_{i(jk)} - T_{j(ik)}$, which are not subject to an identification, but obey the constraint $T_{[xy]z}+T_{[yz]x}+T_{[zx]y}=0$.

\subsection{$\phi$-theory}
There is a $U(1)$ variable $e^{i\phi}$ at each site of the lattice. The action is
\ie\label{XYplaque3p1}
\beta_0 \sum_{\tau\text{-link}} [1-\cos(\Delta_\tau \phi)] + \beta \sum_{i<j}\sum_{ij\text{-plaq}} [1-\cos(\Delta_i \Delta_j \phi)]~,
\fe
where $\phi$ is circle-valued. At large $\beta_0$, $\beta$, we can approximate the action with the Villain action
\ie
\frac{\beta_0}{2} \sum_{\tau\text{-link}} (\Delta_\tau \phi - 2\pi n_\tau)^2 + \frac{\beta}{2} \sum_{i<j}\sum_{ij\text{-plaq}} (\Delta_i \Delta_j \phi - 2\pi n_{ij})^2~,
\fe
where $\phi$ is real and $n_\tau$ and $n_{ij}$ are integer-valued fields on $\tau$-links and $ij$-plaquettes, respectively. There is an integer gauge symmetry
\ie\label{eq:phi-gauge-symmetry}
\phi \sim \phi + 2\pi p~,\qquad n_\tau \sim n_\tau + \Delta_\tau p~,\qquad n_{ij} \sim n_{ij} + \Delta_i \Delta_j p~,
\fe
where $p$ is an integer-valued gauge parameter on the sites. We can interpret $(n_\tau,n_{ij})$ as $\mathbb Z$ tensor gauge fields that make $\phi$ compact.

Next, we suppress the ``vortices'' by modifying the Villain action as
\ie\label{3+1d-phi-modVill-action}
&  \frac{\beta_0}{2} \sum_{\tau\text{-link}} (\Delta_\tau \phi - 2\pi n_\tau)^2 + \frac{\beta}{2} \sum_{i<j}\sum_{ij\text{-plaq}} (\Delta_i \Delta_j \phi - 2\pi n_{ij})^2
\\
&+ i \sum_{i<j} \sum_{\tau ij\text{-cube}} \hat A^{ij} (\Delta_\tau n_{ij} - \Delta_i \Delta_j n_\tau)
-i \sum_{\text{cyclic}\atop i,j,k} \sum_{xyz\text{-cube}}\hat A_\tau^{[ij]k} (\Delta_i n_{jk} - \Delta_j n_{ik})~,
\fe
where $\hat A_\tau^{[ij]k}$ and $\hat A^{ij}$ are real-valued fields on dual $\tau$-links and dual $k$-links respectively. They are Lagrange multipliers that impose the flatness constraint of $(n_\tau,n_{ij})$. They have their own gauge symmetry
\ie
&\hat A_\tau^{[ij]k} \sim \hat A_\tau^{[ij]k} + \Delta_\tau \hat \alpha^{[ij]k} + 2\pi \hat q^{[ij]k}_\tau~,
\\
&\hat A^{ij} \sim \hat A^{ij} + \Delta_k \hat\alpha^{k(ij)} + 2\pi \hat q^{ij}~.
\fe
Here $\hat \alpha^{[ij]k}$ are real-valued fields on the dual sites, while $\hat q^{[ij]k}_\tau$ and $\hat q^{ij}$ are integers on the dual $\tau$-links and the dual $k$-links, respectively.

Following similar steps in Section \ref{sec:2+1phicontinuum}, we can integrate out the real fields $\hat A_\tau^{[ij]k} ,\hat A^{ij}$ and gauge fix most of the integer fields to be zero.
In this gauge choice, the continuum limit of this modified Villain model is recognized as the 3+1d $\phi$-theory of \cite{paper2}.
See also \cite{Slagle:2017wrc,You:2018zhj,Radicevic:2019vyb,Gromov:2020yoc} for related discussions on this theory. Moreover, the modified Villain model has a $U(1)$ $(\mathbf 1, \mathbf 3')$ momentum dipole symmetry and a $U(1)$ $(\mathbf 3', \mathbf 1)$ winding dipole symmetry, which are the same as in the continuum 3+1d $\phi$-theory.

Alternatively, we can apply the Poisson resummation formula \eqref{Possonresummationi} to dualize the modified Villain action \eqref{3+1d-phi-modVill-action} to
\ie\label{3+1d-hatA-modVill-action}
 & \frac{1}{2(2\pi)^2\beta}  \sum_{\text{cyclic}\atop i,j,k} ~ \sum_{\text{dual }\tau k\text{-plaq}} (\Delta_\tau \hat A^{ij} - \Delta_k \hat A_\tau^{k(ij)} - 2\pi \hat n^{ij}_\tau)^2
\\
& + \frac{1}{2(2\pi)^2\beta_0} \sum_{\text{dual }xyz\text{-cube}} \left( \sum_{i<j}\Delta_i \Delta_j \hat A^{ij} - 2\pi \hat n \right)^2 - i \sum_\text{site} \phi \left(\Delta_\tau \hat n - \sum_{i<j}\Delta_i \Delta_j \hat n^{ij}_\tau \right)~,
\fe
where $\hat n^{ij}_\tau$ and $\hat n$ are integer-valued fields on the dual $\tau k$-plaquettes (or $ij$-plaquettes) and the dual hypercubes (or sites) respectively. We interpret $(\hat n^{ij}_\tau,\hat n)$ as $\mathbb Z$ gauge fields that make $(\hat A_\tau^{k(ij)},\hat A^{ij})$ compact via the gauge symmetry\footnote{$(\hat n^{ij}_\tau,\hat n)$ is the $\mathbb Z$ version of $(\hat C^{ij}_\tau,\hat C)$ of \cite{Gorantla:2020xap}.}
\ie\label{eq:hatA-gauge-symmetry}
&\hat A_\tau^{k(ij)} \sim \hat A_\tau^{k(ij)} + \Delta_\tau \hat \alpha^{k(ij)} + 2\pi \hat q^{k(ij)}_\tau~,
\\
&\hat A^{ij} \sim \hat A^{ij} + \Delta_k \hat\alpha^{k(ij)} + 2\pi \hat q^{ij}~,
\\
&\hat n^{ij}_\tau \sim \hat n^{ij}_\tau + \Delta_\tau \hat q^{ij} - \Delta_k \hat q_\tau^{k(ij)}~,
\\
&\hat n \sim \hat n + \sum_{i<j} \Delta_i \Delta_j \hat q^{ij}~.
\fe
The Lagrange multiplier $\phi$ imposes the flatness constraint of $(\hat n^{ij}_\tau,\hat n)$.

Once again, following similar steps in Section \ref{sec:2+1phicontinuum}, we can integrate out the real field $\phi$ and gauge fix most of the integer fields to be zero.
In this gauge choice, the continuum limit of this modified Villain model is recognized as the 3+1d $\hat A$-theory of \cite{paper2} (see also \cite{Slagle:2017wrc,Radicevic:2019vyb}).
Moreover, the modified Villain model has a $U(1)$ $(\mathbf 3', \mathbf 1)$ electric dipole symmetry and a $U(1)$ $(\mathbf 1, \mathbf 3')$ magnetic dipole symmetry, which are the same as in the continuum 3+1d $\hat A$-theory. The duality maps the momentum (winding) dipole symmetry of $\phi$-theory to the magnetic (electric) dipole symmetry of the $\hat A$-theory, exactly like in the continuum theories.

In conclusion, the modified Villain action \eqref{3+1d-phi-modVill-action} has the same continuum limit as the XY-plaquette action \eqref{XYplaque3p1}.  It has all the properties of the continuum $\phi$-theory of \cite{paper2} including the emergent winding symmetry and the duality to the $\hat A$-theory.  It is straightforward to check that the analysis of the singular configurations and the spectrum of charged states of the continuum theory are regularized properly by this modified Villain lattice action.

\subsection{$A$-theory}
There are $U(1)$ variables $e^{i A_\tau}$ and $e^{iA_{ij}}$ on the $\tau$-links and the $ij$-plaquettes of the lattice, respectively. The action is
\ie\label{latticeAthr}
\gamma_0 \sum_{i<j} \sum_{\tau ij\text{-cube}} [1-\cos(\Delta_\tau A_{ij} - \Delta_i \Delta_j A_\tau)] + {\gamma} \sum_{xyz\text{-cube}}~ \sum_{\text{cyclic}\atop i,j,k}  [1-\cos(\Delta_i A_{jk} - \Delta_j A_{ik})]~,
\fe
where $(A_\tau,A_{ij})$ are circle-valued.  This action has a tensor gauge symmetry
\ie\label{eq:A-gauge-symemtryr}
&e^{iA_\tau} \sim e^{iA_\tau + i\Delta_\tau \alpha}~,
\\
&e^{iA_{ij} }\sim e^{iA_{ij} + i\Delta_i \Delta_j \alpha} ~,
\fe
with circle valued $\alpha$ at the sites.

At large $\gamma_0,\gamma$, we can approximate the action, \`{a} la Villain, as
\ie\label{threeAV}
\frac{\gamma_0}{2}\sum_{i<j} \sum_{\tau ij\text{-cube}} (\Delta_\tau A_{ij} - \Delta_i \Delta_j A_\tau - 2\pi n_{\tau ij})^2 + \frac{\gamma}{2} \sum_{xyz\text{-cube}}~ \sum_{\text{cyclic}\atop i,j,k}  (\Delta_i A_{jk} - \Delta_j A_{ik} - 2\pi n_{[ij]k})^2~,
\fe
where now $(A_\tau,A_{ij})$ are real and $n_{\tau ij}$ and $n_{[ij]k}$ are integer-valued fields on the $\tau ij$-cubes and the $xyz$-cubes respectively. The gauge symmetry \eqref{eq:A-gauge-symemtryr} is now replaced with
\ie\label{eq:A-gauge-symemtry}
&A_\tau \sim A_\tau + \Delta_\tau \alpha + 2\pi q_\tau~,
\\
&A_{ij} \sim A_{ij} + \Delta_i \Delta_j \alpha + 2\pi q_{ij}~,
\\
&n_{\tau ij} \sim n_{\tau ij} + \Delta_\tau q_{ij} - \Delta_i \Delta_j q_\tau~,
\\
&n_{[ij]k} \sim n_{[ij]k} + \Delta_i q_{jk} - \Delta_j q_{ik}~.
\fe
Here $\alpha$ is a real-valued field on the sites, while $q_\tau$ and $q_{ij}$ are integer-valued fields on the $\tau$-links and the $ij$-plaquettes, respectively. We interpret $(n_{\tau ij},n_{[ij]k})$ as the $\mathbb Z$ gauge fields that make $(A_\tau,A_{ij})$ compact.\footnote{$(n_{\tau ij},n_{[ij]k})$ is the $\mathbb Z$ version of $(C^{ij}_\tau,C^{[ij]k})$ of \cite{Gorantla:2020xap}.}

Next, we suppress the ``vortices'' by modifying the Villain action as
\ie\label{3+1d-A-modVill-action}
& \frac{\gamma_0}{2}\sum_{i<j} \sum_{\tau ij\text{-cube}} (\Delta_\tau A_{ij} - \Delta_i \Delta_j A_\tau - 2\pi n_{\tau ij})^2 + \frac{\gamma}{2} \sum_{xyz\text{-cube}}~  \sum_{\text{cyclic}\atop i,j,k} (\Delta_i A_{jk} - \Delta_j A_{ik} - 2\pi n_{[ij]k})^2
\\
&\quad +i  \sum_\text{dual site}~ \sum_{\text{cyclic}\atop i,j,k} \hat \phi^{[ij]k} (\Delta_\tau n_{[ij]k} - \Delta_i n_{\tau jk} + \Delta_j n_{\tau ik})~,
\fe
where $\hat \phi^{[ij]k}$ is a real-valued field on the dual sites of the lattice. It is a Lagrange multiplier that imposes the flatness constraint of $(n_{\tau ij},n_{[ij]k})$, and it has a gauge symmetry
\ie
\hat \phi^{[ij]k} \sim \hat \phi^{[ij]k} + 2\pi \hat p^{[ij]k}~,
\fe
where $\hat p^{[ij]k}$ is an integer-valued field on the dual sites.

Following similar steps in Section \ref{sec:2+1phicontinuum}, we can integrate out the real fields $\hat \phi^{[ij]k}$ and gauge fix most of the integer fields to be zero.
In this gauge choice, the continuum limit of this modified Villain model is recognized as the 3+1d $A$-theory of \cite{paper2}.
See also \cite{Xu2008,Slagle:2017wrc,Bulmash:2018lid,Ma:2018nhd,You:2018zhj,Radicevic:2019vyb} for related discussions on this theory. Moreover, the modified Villain model has a $U(1)$ $(\mathbf 3', \mathbf 2)$ electric tensor symmetry and a $U(1)$ $(\mathbf 2, \mathbf 3')$ magnetic tensor symmetry, which are the same as in the continuum 3+1d $A$-theory.

Alternatively, we can apply the Poisson resummation formula \eqref{Possonresummationi} to dualize the modified Villain action \eqref{3+1d-A-modVill-action} to
\ie\label{3+1d-hatphi-modVill-action}
& \frac{1}{6(2\pi)^2\gamma}\sum_{\text{dual }\tau\text{-link}}~ \sum_{\text{cyclic}\atop i,j,k}  (\Delta_\tau \hat \phi^{k(ij)} - 2\pi \hat n^{k(ij)}_\tau)^2
+ \frac{1}{2(2\pi)^2\gamma_0} \sum_{\text{cyclic}\atop i,j,k} ~\sum_{\text{dual }k\text{-link}} (\Delta_k \hat \phi^{k(ij)} - 2\pi \hat n^{ij})^2
\\
& \quad + i  \sum_{\text{cyclic}\atop i,j,k} ~\sum_{ij\text{-plaq}} A_{ij} (\Delta_\tau \hat n^{ij} - \Delta_k \hat n_\tau^{k(ij)}) + i \sum_{\tau\text{-link}} A_\tau \sum_{i<j} \Delta_i \Delta_j \hat n^{ij}~,
\fe
where $\hat n^{k(ij)}_\tau$ and $\hat n^{ij}$ are integer-valued fields on the dual $\tau$-links and the dual $k$-links respectively. There is a gauge symmetry
\ie\label{eq:hatphi-gauge-symmetry}
&\hat \phi^{k(ij)} \sim \hat \phi^{k(ij)} + 2\pi \hat p^{k(ij)}~,
\\
&\hat n_\tau^{k(ij)} \sim \hat n_\tau^{k(ij)} + \Delta_\tau \hat p^{k(ij)}~,
\\
&\hat n^{ij} \sim \hat n^{ij} + \Delta_k \hat p^{k(ij)}~.
\fe
We interpret $(\hat n^{k(ij)}_\tau,\hat n^{ij})$ as $\mathbb Z$ gauge fields that make $\hat\phi^{k(ij)}$ compact. The Lagrange multipliers $(A_\tau,A_{ij})$ impose the flatness constraint of $(\hat n^{k(ij)}_\tau,\hat n^{ij})$. The dual action \eqref{3+1d-hatphi-modVill-action} is the modified Villain action of the $\hat \phi$-theory of \cite{paper2}.

Once again, following similar steps in Section \ref{sec:2+1phicontinuum}, we can integrate out the real fields $(A_\tau,A_{ij})$ and gauge fix most of the integer fields to be zero.
In this gauge choice, the continuum limit of this modified Villain model is recognized as the 3+1d $\hat \phi$-theory of \cite{paper2}. Moreover, the modified Villain model has a $U(1)$ $(\mathbf 2, \mathbf 3')$ momentum tensor symmetry and a $U(1)$ $(\mathbf 3', \mathbf 2)$ winding tensor symmetry, which are the same as in the continuum 3+1d $\hat \phi$-theory. The duality maps the electric (magnetic) tensor symmetry of the $A$-theory to the winding (momentum) tensor symmetry of $\hat \phi$-theory, exactly like in the continuum theories.

To summarize, the lattice $A$-theory \eqref{latticeAthr} and the modified Villain action  \eqref{3+1d-A-modVill-action} flow to the same continuum theory -- the continuum $A$-theory.  The modified Villain action has all the features of the continuum theory.  It has a magnetic symmetry and it is dual to the $\hat\phi$ theory.  It gives a rigorous presentation of the analysis of singular field configurations and the spectrum of charged states found in \cite{paper2}.

\subsection{X-cube model}

In this subsection,
we will start with the X-cube model in its Hamiltonian formalism and deform it to a  modified Villain lattice model.
The latter takes the form of  a $BF$-type action, which admits two equivalent presentations.
The first one, which we call the integer $BF$-action, uses  only the integer fields, while the second one, which we call the real $BF$-action, uses both real and integer fields.
The real $BF$-action is  naturally connected to the continuum $\bZ_N$ tensor gauge theory of \cite{Slagle:2017wrc,paper3}.

\subsubsection{Review of the Hamiltonian formulation}

We start with the Hamiltonian formulation of the X-cube model. On a periodic 3d lattice, there is a $\mathbb{Z}_N$ variable $U$ and its conjugate variable $V$ on each link. They obey $UV=e^{2\pi i/N}VU$. We label the sites by integers $\hat s=(\hat x,\hat y,\hat z)$ and label the links, the plaquette and the cubes using the coordinates of their centers. The Hamiltonian of the X-cube model is \cite{Vijay:2016phm}
\ie\label{Xcube}
&H=-\beta_1\sum_{\text{site}}(G_{\hat s, [yz]x}+G_{\hat s, [zx]y}+G_{\hat s, [xy]z})-\beta_2\sum_{\text{cube}}L_{\hat c}+c.c.~,
\\
&G_{\hat s, [yz]x}=
V_{\hat s+(0,\frac{1}{2},0)}
V_{\hat s+(0,0,\frac{1}{2})}^\dagger
V^\dagger_{\hat s-(0,\frac{1}{2},0)}
V_{\hat s-(0,0,\frac{1}{2})}~,
\\
&G_{\hat s, [zx]y}=
V^\dagger_{\hat s+(\frac{1}{2},0,0)}
V_{\hat s+(0,0,\frac{1}{2})}
V_{\hat s-(\frac{1}{2},0,0)}
V_{\hat s-(0,0,\frac{1}{2})}^\dagger~,
\\
&G_{\hat s, [xy]z}=
V_{\hat s+(\frac{1}{2},0,0)}
V^\dagger_{\hat s+(0,\frac{1}{2},0)}
V^\dagger_{\hat s-(\frac{1}{2},0,0)}
V_{\hat s-(0,\frac{1}{2},0)}~,
\\
&L_{\hat c}=
U_{\hat c+(\frac{1}{2},\frac{1}{2},0)}
U_{\hat c+(-\frac{1}{2},\frac{1}{2},0)}^\dagger
U_{\hat c+(\frac{1}{2},-\frac{1}{2},0)}^\dagger
U_{\hat c-(\frac{1}{2},\frac{1}{2},0)}
\\
&\qquad\
U_{\hat c+(0,\frac{1}{2},\frac{1}{2})}
U_{\hat c+(0,-\frac{1}{2},\frac{1}{2})}^\dagger
U_{\hat c+(0,\frac{1}{2},-\frac{1}{2})}^\dagger
U_{\hat c-(0,\frac{1}{2},\frac{1}{2})}
\\
&\qquad\
U_{\hat c+(\frac{1}{2},0,\frac{1}{2})}
U_{\hat c+(-\frac{1}{2},0,\frac{1}{2})}^\dagger
U_{\hat c+(\frac{1}{2},0,-\frac{1}{2})}^\dagger
U_{\hat c-(\frac{1}{2},0,\frac{1}{2})}~.
\fe
All the terms in the Hamiltonian commute with each other. The operators $G_{\hat s, [ij]k}$ are in the $\mathbf{2}$ of $S_4$ and satisfy $G_{\hat s, [yz]x}G_{\hat s, [zx]y}G_{\hat s, [xy]z}=1$.

The ground states satisfy $G_{\hat s, [ij]k}=L_{\hat c}=1$ for all $\hat s, \hat c$. There are dynamical excitations that violate only $L_{\hat c}=1$ at a cube. Such excitations cannot move so they are fractons. There are also dynamical excitations that violate only $G_{\hat s, [yz]x}=G_{\hat s, [zx]y}=1$ at a site. Such excitations can only move along the $z$ direction so they are $z$-lineons. Similarly, there are $x$-lineons and $y$-lineons that can only move along the $x$ and $y$ direction, respectively. Because of the relation $G_{\hat s, [yz]x}G_{\hat s, [zx]y}G_{\hat s, [xy]z}=1$, an $x$-lineon, a $y$-lineon and a $z$-lineon can annihilate to the vacuum when they meet at the same point.

The X-cube model has a faithful $\mathbb{Z}_N$ $(\mathbf 3',\mathbf 2)$ tensor symmetry and a faithful $\mathbb{Z}_N$ $(\mathbf 3',\mathbf 1)$ dipole symmetry.\footnote{To clarify the terminology, recall that each symmetry operator is associated with a geometrical object $\cal C$.  According to \cite{Qi:2020jrf}, if the action of the operator depends only on the topology of $\cal C$, the symmetry is not faithful, while if it depends also on its geometry, the symmetry is faithful. For example, the non-relativistic $q$-form symmetry  of \cite{Seiberg:2019vrp} is faithful, while the relativistic $q$-form symmetry of \cite{Gaiotto:2014kfa} is not faithful. In \cite{paper3}, the faithful symmetry was referred to as ``unconstrained'' and the unfaithful symmetry was referred to as ``constrained.''} A typical symmetry operator of the faithful $\mathbb{Z}_N$ $(\mathbf 3',\mathbf 2)$ tensor symmetry is the line operator
$\prod_{z\text{-link: fixed }\hat y,\hat z} U$.  And there are similar lines along other directions.
A typical symmetry operator of the faithful $\mathbb{Z}_N$ $(\mathbf 3',\mathbf 1)$ dipole symmetry is $\prod_{\mathcal C^{xy}} V$
where $\mathcal C^{xy}$ is a closed curve along the dual links at fixed $\hat z_0$. Similarly, there are other symmetry operators on the other planes.

We are interested in the $\beta_1,\beta_2\rightarrow\infty$ limit of the model. In this limit, $G_{\hat s, [ij]k}=L_{\hat c}=1$ for all $\hat s, \hat c$ and the Hilbert space is restricted to the ground states.

\subsubsection{Integer $BF$-action}

We now formulate the X-cube model in the $\beta_1,\beta_2\rightarrow\infty$ limit in the Lagrangian formalism. We put the model on a periodic 4d Euclidean lattice. For each $k$-link, we introduce an integer-valued field $\hat m^{ij}$ with $i\neq j\neq k$ for the $\mathbb{Z}_N$ variable $U=\exp(\frac{2\pi i \hat m^{ij}}{N} )$. For each dual $ij$-plaquette, we introduce an integer-valued field $m_{ij}$ for the conjugate $\mathbb{Z}_N$ variable $V=\exp(\frac{2\pi i m_{ij}}{N} )$.

Next, we  introduce Lagrange multiplier fields to impose the constraints $G_{\hat s, [ij]k}=L_{\hat c}=1$.
On each dual $\tau$-link (or $xyz$-cube), we introduce an integer-valued field $m_\tau$ to impose $L_{\hat c}=1$ as a constraint. On each $\tau$-link, we introduce three integer-valued fields $\hat m_\tau^{[ij]k}$ to impose $G_{\hat s, [ij]k}=1$ as constraints.
Since $G_{\hat s, [yz]x}G_{\hat s, [zx]y}G_{\hat s, [xy]z}=1$, one combination of $\hat m_\tau^{[ij]k}$ decouples and therefore $\hat m_\tau^{[ij]k}$ has a gauge symmetry. Below, we will instead work with the combinations $\hat m_\tau^{k(ij)} = \hat m_\tau^{[ki]j} - \hat m_\tau^{[jk]i}$, which are not subject to any gauge symmetry, but are constrained to satisfy $\hat m_\tau^{x(yz)}+\hat m_\tau^{y(zx)}+\hat m_\tau^{z(xy)}=0$.

In terms of these integer fields, the Euclidean lattice action for the low-energy limit of the X-cube model is
\ie\label{3+1d-ZN-modVill-action}
&\frac{2\pi i}{N} \sum_{\text{cyclic}\atop i,j,k} ~\sum_{\tau k\text{-plaq}}m_{ij}\left(\Delta_\tau\hat m^{ij}-\Delta_k \hat m_\tau^{k(ij)}\right)
+\frac{2\pi i}{N} \sum_{xyz\text{-cube}} m_\tau\left(\sum_{i<j}\Delta_i \Delta_j \hat m^{ij}\right)~.
\fe
There are gauge symmetries:
\ie
&m_\tau\sim m_\tau +\Delta_\tau \ell + Nq_\tau~,
\\
&m_{ij}\sim m_{ij}+\Delta_i\Delta_j\ell+Nq_{ij}~,
\\
&\hat m_\tau^{k(ij)}\sim\hat m_\tau^{k(ij)}+\Delta_\tau \hat \ell^{k(ij)}
+N\hat q_\tau^{k(ij)}~,
\\
&\hat m^{ij}\sim\hat m^{ij}+\Delta_k \hat \ell^{k(ij)}+N\hat q^{ij}~,
\fe
where $\ell$, $\hat\ell^{k(ij)}$, $q_\tau$, $q_{ij}$,
$\hat q_\tau^{k(ij)}$ and $\hat q^{ij}$ are integer-valued fields on the dual sites, the sites, the dual $\tau$-links, the dual $ij$-plaquettes, the $\tau$-links,
and the $k$-links, respectively.
We will refer to this presentation of the model as the integer $BF$-action.
This is analogous to the presentation \eqref{eq:ZNtopoaction} for the topological lattice $\bZ_N$ gauge theory reviewed in Appendix \ref{pformgau}  and the presentation \eqref{2+1d-ZN-tensor-Vill-action} of the 2+1d tensor $\bZ_N$ tensor gauge theory.

The fields $(\hat m_{\tau}^{k(ij)},\hat m^{ij})$ and $(m_\tau,m_{ij})$ pair up into two integer-valued tensor gauge fields. Comparing with \eqref{3+1d-hatA-modVill-action} and \eqref{3+1d-A-modVill-action}, we can interpret \eqref{3+1d-ZN-modVill-action} as the $\mathbb{Z}_N$ lattice tensor gauge theory of the $\hat A$ gauge field or the $ A$ gauge field.

In this Lagrangian,  there are no dynamical fractons and lineons. Instead, charged particles become defects of probe fractons and lineons.
The probe fracton defect is
\ie\label{fracton}
W^\tau(\hat x,\hat y,\hat z)=\exp\left[ \frac{2\pi i}{N} \sum_{\text{dual }\tau\text{-link: fixed }\hat x,\hat y,\hat z} m_\tau\right]~,
\fe
and the probe $z$-lineon defect is
\ie\label{z-lineon}
\hat W^z(\hat x,\hat y,\mathcal C^z)=\exp\left[ \frac{2\pi i}{N} \sum_{\tau\text{-link}\in \mathcal C^z} \hat m_\tau^{z(xy)} + \frac{2\pi i}{N} \sum_{z\text{-link}\in \mathcal C^z}\hat m^{xy}\right]~,
\fe
where $\mathcal C^z$ is a curve along the $\tau$- and $z$-links in the $\tau z$-plane at fixed $\hat x$ and $\hat y$. The $x$- and $y$-lineons are defined similarly.

The $\mathbb{Z}_N$ lattice tensor gauge theory has a $\mathbb{Z}_N$ $(\mathbf 3',\mathbf 2)$ tensor symmetry and a $\mathbb{Z}_N$ $(\mathbf 3',\mathbf 1)$ dipole symmetry. The $\mathbb{Z}_N$ $(\mathbf 3',\mathbf 2)$ tensor symmetry is generated by the line operator of \eqref{z-lineon} along a closed curve $\mathcal{C}^z$ and other similar line operators on the $\tau x$- and $\tau y$-plane. These symmetry operators are constrained by the flatness condition on $\hat m^{ij}$. So, the $\mathbb{Z}_N$ $(\mathbf 3',\mathbf 2)$ tensor symmetry is unfaithful (in the sense of \cite{Qi:2020jrf}). The charged observables are the probe fracton defect \eqref{fracton} and the Wilson observable
\ie\label{Wilsonfracton}
&W^{xy}(\hat z,\mathcal C^{xy}) = \exp\left[ \frac{2\pi i}{N} \left(\sum_{\text{dual }xz\text{-plaq}\in\mathcal C^{xy}} m_{xz} +  \sum_{\text{dual }yz\text{-plaq}\in\mathcal C^{xy}} m_{yz} + \sum_{\text{dual }\tau z\text{-plaq}\in\mathcal C^{xy}} \Delta_z m_\tau\right) \right]~,
\fe
where $\mathcal C^{xy}$ is a closed strip along the $xz$-, $yz$- and $\tau z$-plaquettes at a fixed $\hat z$. Similarly, there are other charged Wilson observables $ W^{yz}(\hat x,\mathcal C^{yz})$ and $W^{zx}(\hat y,\mathcal C^{zx})$. The $\mathbb{Z}_N$ $(\mathbf 3',\mathbf 1)$ dipole symmetry is generated by the line operator \eqref{fracton},  \eqref{Wilsonfracton} and similar lines operators at fixed $\hat x$ or $\hat y$. These symmetry operators are quasi-topological, i.e., they are invariant under small deformation of $\mathcal{C}^{xy}$ on the $\tau x y$-volume. So, the $\mathbb{Z}_N$ $(\mathbf 3',\mathbf 1)$ dipole symmetry is unfaithful (in the sense of \cite{Qi:2020jrf}). The charge operators are \eqref{z-lineon} and similar operators on the other planes.

\subsubsection{Real $BF$-action and the continuum limit}

As in Section \ref{sec:2+1realBF}, we discuss another presentation of this theory, which is closer to the continuum action.

Starting from the integer $BF$-action \eqref{3+1d-ZN-modVill-action}, we replace the integer-valued fields $(m_\tau,m_{ij})$ and $(\hat m^{k(ij)}_\tau,\hat m^{ij})$ with real-valued fields $(A_\tau,A_{ij})$ and $(\hat A^{k(ij)}_\tau,\hat A^{ij})$.  We constrain them to be integer-valued using Lagrange multiplier fields $(\hat n^{ij}_\tau,\hat n)$ and $(n_{\tau ij},n_{[ij]k})$ . Furthermore, since the gauge fields $(A_\tau,A_{ij})$ and $(\hat A^{k(ij)}_\tau,\hat A^{ij})$ have real-valued gauge symmetries, we introduce Stueckelberg fields $\phi$ and $\hat \phi^{[ij]k}$ for their gauge symmetries.   We end up with the action
\ie\label{3+1d-ZN-modVill-real-action}
&\frac{ iN}{2\pi}\sum_{\tau k\text{-plaq}}A_{ij}\left(\Delta_\tau\hat A^{ij}-\Delta_k\hat A_\tau^{k(ij)}-2\pi \hat n_{\tau}^{ij}\right)
+\frac{iN}{2\pi}\sum_{xyz\text{-cube}} A_\tau \left(\Delta_i \Delta_j \hat A^{ij}-2\pi \hat n\right)
\\
&+iN\sum_{ij\text{-plaq}}\hat A^{ij}n_{\tau ij}-{iN}\sum_{\tau\text{-link}} \hat A^{[ij]k}_\tau n_{[ij]k}-i \sum_{\text{dual site}}\phi\left(\Delta_\tau\hat n- \Delta_i \Delta_j \hat n^{ij}_\tau \right)
\\
&-i \sum_\text{site}\hat \phi^{[ij]k} (\Delta_\tau n_{[ij]k} - \Delta_i n_{\tau jk} + \Delta_j n_{\tau ik})
~.
\fe
(To simplify this particular expression and \eqref{3+1d-ZN-tensor-HiggsAction}, we use the convention that repeated indices $i,j$ and  $i,j,k$ are  summed over cyclically.) Here $\phi$, $\hat \phi^{[ij]k}$, $A_\tau$, $A_{ij}$, $\hat A_\tau^{[ij]k}$ and $\hat A^{ij}$ are real-valued fields on dual sites, sites, dual $\tau$-links, dual $ij$-plaquettes, $\tau$-links and $k$-links, respectively, and $n_{\tau ij}$, $n_{[ij]k}$, $\hat n^{ij}_\tau$ and $\hat n$ are integer-valued fields on the dual $\tau ij$-cubes, the dual $xyz$-cubes, the $\tau k$-plaquettes, and the $xyz$-cubes, respectively.   We will refer to this presentation as the real $BF$-action, which uses both the real and integer fields.

These fields have the same gauge symmetries as in \eqref{eq:phi-gauge-symmetry}, \eqref{eq:hatA-gauge-symmetry} \eqref{eq:A-gauge-symemtry}, \eqref{eq:hatphi-gauge-symmetry}
except that the $\alpha$ and $\hat\alpha^{[ij]k}$ gauge symmetry also acts on $\phi$ and $\hat \phi^{[ij]k}$ as
\ie\label{3+1d-ZN-modVill-real-gaugesym}
&\phi\sim \phi+N\alpha~,
\\
&\hat \phi^{[ij]k}\sim \hat \phi^{[ij]k}+N\hat\alpha^{[ij]k}~.
\fe

As a check, summing over $(\hat n^{ij}_\tau,\hat n)$ and $(n_{\tau ij},n_{[ij]k})$ in \eqref{3+1d-ZN-modVill-real-action} constrains
\ie &\left(A_{\tau}-\frac{1}{N}\Delta_\tau\phi,A_{ij}-\frac{1}{N}\Delta_i\Delta_j\phi\right)=\frac{2\pi}{N}(m_\tau,m_{ij}) ~,
\\
&\left(\hat A_\tau^{k(ij)}-\frac{1}{N}\Delta_\tau \hat \phi^{k(ij)},\hat A^{ij}-\frac{1}{N}\Delta_k\hat \phi^{k(ij)}\right)=\frac{2\pi}{N}(\hat m_\tau^{k(ij)},\hat m^{ij})~.
\fe
Substituting them back to the action leads to \eqref{3+1d-ZN-modVill-action}.

The real $BF$-action \eqref{3+1d-ZN-modVill-real-action} can also be derived through Higgsing the $U(1)$ tensor gauge theory \eqref{3+1d-A-modVill-action} to a $\mathbb{Z}_N$ theory using the field $\phi$ in \eqref{3+1d-phi-modVill-action}. The Higgs action is
\ie\label{3+1d-ZN-tensor-HiggsAction}
&  \frac{i}{2\pi}\sum_{\tau\text{-link}} \hat B(\Delta_\tau \phi-NA_\tau - 2\pi n_\tau) +  \frac{i}{2\pi}\sum_{ij\text{-plaq}} \hat E^{ij}(\Delta_i \Delta_j \phi-NA_{ij} - 2\pi n_{ij})
\\
&- i  \sum_{\tau ij\text{-cube}} \hat A^{ij} (\Delta_\tau n_{ij} - \Delta_i \Delta_j n_\tau- N n_{\tau ij})
+i  \sum_{xyz\text{-cube}}\hat A_\tau^{[ij]k} (\Delta_i n_{jk} - \Delta_j n_{ik}- N n_{[ij]k})
\\
& -i  \sum_\text{dual site}~  \hat \phi^{[ij]k} (\Delta_\tau n_{[ij]k} - \Delta_i n_{\tau jk} + \Delta_j n_{\tau ik})~,
\fe
where $\hat B$ and $\hat E^{ij}$ are real-valued fields on the $\tau$-links and the $ij$-plaquattes, respectively. These fields have the same gauge symmetries as in \eqref{eq:phi-gauge-symmetry}, \eqref{eq:hatA-gauge-symmetry}, \eqref{eq:A-gauge-symemtry}, \eqref{eq:hatphi-gauge-symmetry}, and \eqref{3+1d-ZN-modVill-real-gaugesym}. In addition, the fields $(n_\tau,n_{ij})$ also transform under the $(q_\tau,q_{ij})$ gauge symmetry
\ie
&n_\tau\sim n_\tau -Nq_\tau~,
\\
&n_{ij}\sim n_{ij}-Nq_{ij}~.
\fe
Summing over the integer-valued fields $(n_\tau, n_{ij})$ constrains
\ie
\hat B-\sum_{i<j}\Delta_i\Delta_j\hat A^{ij}=-2\pi \hat n~,\quad\hat E^{ij}-\Delta_\tau\hat A^{ij}+\Delta_k\hat A^{k(ij)}_\tau=-2\pi\hat n^{ij}_\tau~,
\fe
where $\hat n$ and $\hat n_\tau^{ij}$ are integer-valued fields. Substituting them back into the action leads to \eqref{3+1d-ZN-modVill-real-action}. Similarly, the real $BF$-action \eqref{3+1d-ZN-modVill-real-action} can also be derived through Higgsing the $U(1)$ tensor gauge theory \eqref{3+1d-hatA-modVill-action} to a $\mathbb{Z}_N$ theory using the field $\hat\phi^{[ij]k}$ in \eqref{3+1d-hatphi-modVill-action}.

Let us discuss a convenient gauge choice for this lattice model.
Following similar steps in Section \ref{sec:2+1realBF} and in Appendix \ref{ZNgauget}, we first integrate out $\phi$ and $\hat \phi^{[ij]k}$, and then gauge fix most of the integers $(n_{\tau ij},n_{[ij]k})$ and $(\hat n^{ij}_\tau,\hat n)$ to zero.
Next, we define new fields that are not single-valued and have transition functions.
In this gauge choice, it is then straightforward to take the continuum limit of the real $BF$-action:
\ie\label{3+1d-ZN-action-contlim}
&\frac{ iN}{2\pi}\int d\tau dxdydz\left[ \sum_{\text{cyclic}\atop i,j,k} A_{ij}\left(\partial_\tau\hat A^{ij}-\partial_k\hat A_\tau^{k(ij)}\right)
+A_\tau \left(\sum_{i<j}\partial_i \partial_j \hat A^{ij}\right)\right]~,
\fe
where we omit the terms that depend on the transition functions of these fields.\footnote{Such boundary terms are necessary in order to make the continuum action \eqref{3+1d-ZN-action-contlim} well-defined.  They played a crucial role in the analysis of \cite{Rudelius:2020kta}.} This is the Euclidean version of the 3+1d $\mathbb{Z}_N$ tensor gauge theory of \cite{Slagle:2017wrc,paper3} which  describes the low-energy limit of the X-cube model.

We conclude that the modified Villain lattice model \eqref{3+1d-ZN-modVill-action}, or equivalently \eqref{3+1d-ZN-modVill-real-action}, flows
to the same continuum field theory  \eqref{3+1d-ZN-action-contlim} as the original X-cube model \eqref{Xcube}.
Conversely, the modified Villain lattice model \eqref{3+1d-ZN-modVill-action}, or equivalently \eqref{3+1d-ZN-modVill-real-action}, gives a rigorous setting for the discussion of the continuum theory \eqref{3+1d-ZN-action-contlim} of \cite{Slagle:2017wrc,paper3}.

\section*{Acknowledgements}

We thank F.~Burnell, A.~Kapustin, Z.~Komargodski, S.~Sachdev, S.~Shenker,  T.~Sulejmanpasic, and A.~Tiwari for helpful discussions and comments. PG was supported by Physics Department of Princeton University.  HTL was supported
by a Centennial Fellowship and a
Charlotte Elizabeth Procter Fellowship from Princeton University and Physics Department of Princeton University. The work of NS was supported in part by DOE grant DE$-$SC0009988.  NS and SHS were also supported by the Simons Collaboration on Ultra-Quantum Matter, which is a grant from the Simons Foundation (651440, NS).
Opinions and conclusions expressed here are those of the authors and do not necessarily reflect the views of funding agencies.

\appendix
\section{Villain formulation of some classic quantum-mechanical systems}\label{sec:VillainQM}

 In this appendix,  we review two classic quantum-mechanical systems.
 The various versions of the theory that we will present and the manipulations of the equations are simple warmup examples for the other models.

\subsection{Particle on a ring}

We start with the quantum mechanics of a particle on a ring parameterized by the periodic coordinate $q\sim q+2\pi$.  This problem is a classic example of the $\theta$-parameter and its effects.  We discuss it using the lattice Villain formulation.

The problem is characterized by the Euclidean continuum action
\ie\label{particler}
S=\oint d\tau \left({1\over 2}(\partial_\tau q)^2 +{i\theta\over 2\pi} \partial_\tau q\right)~
\fe
and we take the circumference of the Euclidean-time circle to be $\ell$.  The $\theta$-parameter is $2\pi$-periodic. (Here, we used the freedom to rescale $\tau$ to set the coefficient of the kinetic term to $1\over 2$.)

This system has a global $U(1)$ symmetry shifting $q$ by a constant.  And for $\theta\in \pi \bZ$, it also has a charge conjugation symmetry $q\to -q$. These two symmetries combine to $O(2)$.   As emphasized in \cite{Gaiotto:2017yup}, for $\theta\in (2\bZ+1)\pi$ there is an 't Hooft anomaly stating that while the operator algebra has an $O(2)$ symmetry, the Hilbert space realizes it projectively.  Related to that, this system has an anomaly in the space of coupling constants \cite{Cordova:2019jnf,Cordova:2019uob}.  We are going to reproduce these results on a Euclidean lattice.

Next, we place this theory on a Euclidean-time lattice with lattice spacing $a$.  We label the sites by $\hat \tau \in \bZ$ such that $\tau=a\hat\tau$ and the total number of sites is $L=\ell/a$.  Then, following the Villain approach, we make the coordinate $q(\hat\tau)$ real-valued and add an integer-valued gauge field on the
links.  The lattice Lagrangian and action are
\ie\label{particlerL}
&{\cal L}={1\over 2a}\big(\Delta q(\hat \tau) -2\pi n(\hat\tau)\big)^2 +{i\theta\over 2\pi} \big(\Delta q(\hat \tau) -2\pi n(\hat\tau)\big)~,\\
&S=\sum_{\hat \tau=0}^{L-1}{\cal L}\\
&\Delta q(\hat \tau) = q(\hat \tau+1)-q(\hat \tau)~.
\fe
This system has a $\bZ$ gauge symmetry
\ie\label{Zgaugeqm}
&q(\hat \tau) \sim q(\hat\tau ) + 2\pi k(\hat \tau)\\
&n(\hat \tau) \sim n(\hat\tau) +\Delta_\tau k(\hat\tau)\\
&k(\hat\tau)\in \bZ~.
\fe
We can replace the Lagrangian in \eqref{particlerL} by
\ie\label{altLaqm}
{\cal L'}={1\over 2a}\big(\Delta q(\hat \tau) -2\pi n(\hat\tau)\big)^2 -i\theta n(\hat \tau)
\fe
without changing the action.  Unlike $\cal L$, the new Lagrangian $\cal L'$ is not gauge invariant under \eqref{Zgaugeqm}.

The main point about \eqref{particlerL} or \eqref{altLaqm} is the description of the $\theta$-term using the gauge field. The integer topological charge of the continuum theory ${1\over 2\pi}\oint d\tau \partial_\tau q$ is described by the Wilson line of $n$.

As in the continuum, the global $U(1)$ symmetry acts by shifting $q$ by a constant.  It is $U(1)$ rather than $\bR$ because its subgroup $\bZ\subset \bR$ is gauged.  The charge conjugation operation $q\to -q$ should be combined with $n\to -n$.  Unless $\theta=0$, it is not a symmetry of the action \eqref{particlerL}.  However, for $\theta\in \pi\bZ$, it is a symmetry of $e^{-S}$.

Let us examine the charge conjugation symmetry more carefully.  Its action is ``on-site.''  However, unless $\theta=0$, it does not leave the Lagrangian $\cal L$ or even the action $S$ in \eqref{particlerL} invariant. It does not even leave the exponential of the Lagrangian $e^{-{\cal L}}$ invariant.  The symmetry is present for $\theta\in \pi\bZ$ because it leaves $e^{-S}$ invariant.  This opens the door for an 't Hooft anomaly associated with this symmetry and to the related anomaly in the space of coupling constants of \cite{Cordova:2019jnf,Cordova:2019uob}.\footnote{Note that  $e^{-{\cal L'}}$ with $\cal L'$ of \eqref{altLaqm} is $O(2)$ invariant for $\theta \in \pi \bZ$, but it is not gauge invariant.  This is common with anomalies.  Using counterterms, we can move the problem around, but we cannot get rid of it.}

This anomaly is exactly as in the continuum discussion of \cite{Gaiotto:2017yup}.  It can be demonstrated by adding to \eqref{particlerL} a classical $U(1)$ gauge field $A$
\ie\label{particlerLg}
&{\cal L}={1\over 2a}\big(\Delta q(\hat \tau) -A(\hat \tau) -2\pi n(\hat\tau)\big)^2 +{i\theta\over 2\pi} \big(\Delta q(\hat \tau)-A(\hat \tau) -2\pi n(\hat\tau)\big)~.
\fe
To see that the gauge symmetry of $A$ is $U(1)$ rather than $\bR$, we note that its gauge symmetry
\ie\label{ZgaugeqmA}
&q(\hat \tau) \sim q(\hat\tau ) +\Lambda(\hat\tau)+ 2\pi k(\hat \tau)\\
&n(\hat \tau) \sim n(\hat\tau) +\Delta_\tau k(\hat\tau)-N(\hat \tau)\\
&A(\hat\tau) \sim A(\hat \tau) + \Delta \Lambda (\hat \tau) +2\pi N(\hat \tau)\\
&k(\hat\tau),N(\hat \tau)\in \bZ~
\fe
includes a $\bZ$ one-form gauge symmetry with the integer gauge parameter $N(\hat \tau)$.  Invariance under this gauge symmetry shows that the $\theta$-term must depend on $A$ even if we use $\cal L'$ of \eqref{altLaqm}.\footnote{
An extreme version of this system is when the lattice has only one site, i.e., $L=1$.  In that case the action becomes
\ie\label{particlerLA}
{1\over 2a}\big(A(\hat \tau) +2\pi n(\hat\tau)\big)^2 -{i\theta\over 2\pi} \big(A(\hat \tau) +2\pi n(\hat\tau)\big)~.
\fe
The global $U(1)$ symmetry is reflected in the fact that action is independent of $q$.  It depends only on the integer dynamical gauge field $n$ and the classical gauge field $A$.  The remaining gauge symmetry is the one-form gauge symmetry
\ie\label{ZgaugeqmAr}
&n(\hat \tau) \sim n(\hat\tau) -N(\hat \tau)\\
&A(\hat\tau) \sim A(\hat \tau) +2\pi N(\hat \tau)\\
&N(\hat \tau)\in \bZ~.
\fe
Again, the anomaly is manifest in \eqref{particlerLA}.}   Now, the charge conjugation symmetry acts also on $A$ and as a result, the $\theta$-term is not invariant under it unless $\theta=0$.  As in \cite{Cordova:2019jnf,Cordova:2019uob}, this also means that there is an anomaly in the $2\pi$-periodicity in $\theta$.

One way to think about this lattice model is the following.  We choose the gauge $n(\hat\tau)=0$ except for $n(0)$.  In this gauge the Wilson line of $n$ is given by $n(0)$, which is gauge invariant.  The remaining gauge symmetry is the identification $q\sim q+2\pi k$ with integer $k$ independent of $\hat \tau$.  It is convenient to redefine $q$ to the nonperiodic (in $\hat \tau$) variable
\ie
&\bar q(\hat\tau)=
\begin{cases}
	q(\hat\tau)\quad\text{for }\hat\tau=1,\cdots,L
	\\
	q( 0)+2\pi n(0)\quad\text{for }\hat\tau=0
\end{cases}~.
\fe
In these variables, after dropping the bar, \eqref{particlerL} becomes
\ie\label{particlerLs}
&{\cal L}={1\over 2a}\big(\Delta q(\hat \tau)\big)^2 +{i\theta\over 2\pi} \Delta q(\hat \tau)~,\\
&S=\sum_{\hat \tau=0}^{L-1}{\cal L}~.
\fe
This can be interpreted as follows.  We have a real-valued field $q$ and we sum over twisted boundary conditions labeled by an integer $n(0)$ such that $q(\hat \tau +L)=q(\hat \tau) - 2\pi n(0)$.

In the form \eqref{particlerLs}, it is easy to take the continuum limit.  We take $a\to 0$, $L\to \infty$ with finite $\ell=La$.  In this limit $q$ becomes smooth and we recover \eqref{particler}.

\subsection{Noncommutative torus}

Next, we review the quantum mechanics of $N$ degenerate ground states using a Euclidean lattice.

In the continuum, the theory can be described using a phase space of two circle-valued coordinates $p,q$ with the Euclidean action
\ie\label{eq:pqdotaction}
\frac{iN}{2\pi}\int d\tau\, p\dot q~.
\fe
(Soon, we will make this action more precise.)  Its quantization leads to $N$ degenerate ground states. These ground states are in the minimal representation of the operator algebra
\ie
&UV=e^{\frac{2\pi i}{N}}VU,
\\
&U=e^{ip},\quad V=e^{iq}~.
\fe

Since $p$ and $q$ are circle-valued, i.e., $p(\tau)\sim p(\tau )+2\pi$ and $q(\tau)\sim q(\tau)+2\pi$, the Lagrangian in \eqref{eq:pqdotaction} is not well defined.   There are several ways to correct it.  One of them involves lifting $q$ and $p$ to be real-valued with transition functions at some reference point $\tau_*$.  Then, we can take the action to be \cite{Bauer:2004nh,Cordova:2019jnf,Cordova:2019uob} (see also \cite{Freed:2006ya,Freed:2006yc,Kapustin:2014gua,Rudelius:2020kta})\footnote{The rigorous mathematical treatment uses  differential cohomology \cite{ChS,De1,DF,HS} (see \cite{Bu,Sc,BNV} and the references therein for modern developments).}
\ie\label{eq:pqdotactionc}
\frac{iN}{2\pi}\int_{\tau_*}^{\tau_*+\ell}d\tau\,p\dot q - i N w_p(\tau_*)q(\tau_*)~,
\fe
where $\ell$ is the period of the Euclidean time and $w_p=\frac{1}{2\pi}[p(\tau_*+\ell)-p(\tau_*)]$ is the winding number of $p$.
Similarly, we define  $w_q=\frac{1}{2\pi}[q(\tau_*+\ell)-q(\tau_*)]$ as the winding number of $q$.
In the path integral, we sum over the integers $w_p$ and $w_q$.
The action is independent of the choice of $\tau_*$, i.e., the choice of trivialization.

Note that as in \eqref{particler}, we could have added to \eqref{eq:pqdotactionc} $\theta$-terms for $p$ and $q$.  However, it is clear that they can be absorbed in shifts of $q$ and $p$ respectively.  Therefore, without loss of generality, we can ignore them.  The same comment applies to the lattice discussion below.

We now discretize the Euclidean time direction and replace it by a periodic lattice with $\tau=a\hat \tau$, $\hat \tau\in\mathbb{Z}$ and periodicity $\hat \tau\sim\hat \tau +L$.  We use the Villain approach and let $q$ and $p$ be real-valued (as opposed to circle-valued) coordinates coupled to $\bZ$ gauge fields $n_q$ and $n_p$.  The action is
\ie\label{eq:Villainpqdot}
&\frac{iN}{2\pi}\sum_{\hat\tau=0}^{L-1} \Big[p(\hat\tau)\big(\Delta q(\hat \tau)-2\pi n_q(\hat\tau)\big)+ 2\pi n_p(\hat \tau) q(\hat\tau)\Big]~,\\
&\Delta  q(\hat\tau)\equiv q(\hat \tau+1)-  q(\hat \tau)~.
\fe
The fields $q,n_p$ naturally live on the lattice sites, while $p,n_q$ naturally live on the links. These fields are subject to gauge symmetries with integer gauge parameters $k_p,k_q$
\ie
&p(\hat \tau)\sim p(\hat \tau)+2\pi k_p(\hat\tau)~,
\\
&q(\hat \tau)\sim q(\hat \tau)+2\pi k_q(\hat\tau)~,
\\
&n_p(\hat \tau)\sim n_p(\hat \tau)+ k_p(\hat\tau)-k_p(\hat\tau-1)~,
\\
&n_q(\hat \tau)\sim n_q(\hat \tau)+ k_q(\hat\tau+1)- k_q(\hat\tau)~.
\fe
Note that the Lagrangian is not gauge invariant.  Even the action is not gauge invariant.  But $e^{-S} $ is gauge invariant.

We can choose the gauge $n_q(\hat\tau)=n_p(\hat{\tau})=0$ except for $n_q(0), n_p(0)$. The action then becomes
\ie\label{gaugecom}
\frac{iN}{2\pi}\sum_{\hat\tau=0}^{L-1} p(\hat\tau)\Delta q(\hat \tau)-iN n_q(0)p(0)+ iN n_p(0) q(0)~.
\fe
There is a residual gauge symmetry:
\ie
&p(\hat \tau)\sim p(\hat \tau)+2\pi ~,
\\
&q(\hat \tau)\sim q(\hat \tau)+2\pi ~.
\fe
To relate the gauge fixed lattice action \eqref{gaugecom} to the continuum action \eqref{eq:pqdotactionc}, we define new variables $\bar p,\bar q$ on the covering space of the periodic lattice:
\ie
&\bar p(\hat\tau)=
\begin{cases}
	 p(\hat\tau)\quad\text{for }\hat\tau=0,\cdots,L-1
	 \\
	 p(\hat\tau)-2\pi n_p(0)\quad\text{for }\hat\tau=L
\end{cases}~,
\\
&\bar q(\hat\tau)=
\begin{cases}
	q(\hat\tau)\quad\text{for }\hat\tau=1,\cdots,L
	\\
	q(0)+2\pi n_q(0)\quad\text{for }\hat\tau=0
\end{cases}~.
\fe
Unlike the single-valued real fields $p,q$, which obey $p(0)=p(L)$, $q(0)=q(L)$, the new real fields $\bar p,\bar q$ are not single-valued on the periodic lattice; they can have non-trivial winding number $w_p=-n_p(0)$, $w_q=-n_q(0)$.
In terms of the new variables, the action becomes
\ie
\frac{iN}{2\pi}\sum_{\hat\tau=0}^{L-1} \bar p(\hat\tau)\Delta\bar q(\hat \tau)- iN w_p \bar q(0)~,
\fe
In the continuum limit, this lattice action  becomes \eqref{eq:pqdotactionc}.

Instead of gauge fixing the integer fields $n_p,n_q$, we can sum over them. This restricts the real-valued fields $p,q$ to $p=\frac{2\pi}{N}m_p$ and $q=\frac{2\pi}{N} m_q$ with integer fields $m_p,m_q$. The action becomes
\ie
\frac{2\pi i}{N}\sum_{\hat\tau=1}^Lm_p(\hat\tau)\Delta m_q(\hat\tau)~,
\fe
with the following gauge symmetry making the integer fields $\mathbb{Z}_N$ variables
\ie
&m_p(\hat\tau)\sim m_p(\hat\tau)+Nk_p(\hat\tau)~,
\\
&m_q(\hat\tau)\sim m_q(\hat\tau)+Nk_q(\hat\tau)~.
\fe

\section{Modified Villain formulation of 2d Euclidean lattice theories without gauge fields}\label{Villainlattia}

In this appendix, we review well-known facts about some lattice models and their Villain formulation.  As in the models in the bulk of the paper, we deform the standard Villain action to another lattice action, which has special properties.  In particular, it has enhanced global symmetries and it exhibits special dualities.  Then, we study other models by deforming this special action.

\subsection{2d Euclidean XY-model}\label{2dXY}

Here we study the two-dimensional Euclidean XY-model on the lattice and in the continuum limit \cite{PhysRevB.16.1217,PhysRevB.17.1340}.

\subsubsection{Lattice models}

We place the theory on a 2d Euclidean periodic lattice, whose sites are labeled by integers $(\hat x, \hat y)\sim (\hat x+L^x, \hat y)\sim (\hat x, \hat y+L^y) $. The dynamical variables are phases  $e^{i\phi}$ at each site of the lattice. The action is
\ie\label{XY-action}
\beta \sum_\text{link} [1- \cos(\Delta_\mu \phi)]~,
\fe
where $\mu=x,y$ labels the directions and $\Delta_x\phi\equiv \phi(\hat x +1,\hat y) - \phi(\hat x, \hat y)$ and $\Delta_y\phi\equiv \phi(\hat x ,\hat y+1)- \phi(\hat x, \hat y)$ are the lattice derivatives.

At large $\beta$, we can approximate the action \eqref{XY-action} by the Villain action \cite{Villain:1974ir}:
\ie\label{XY-Villain-action}
\frac{\beta}{2} \sum_\text{link} (\Delta_\mu \phi - 2\pi n_\mu)^2~.
\fe
Here $\phi$ is a real-valued field and $n_\mu$ is an integer-valued field on the links.  These fields satisfy periodic boundary conditions.

The fact that in the original formulation \eqref{XY-action}, $\phi$ was circle-valued rather than real-valued is related to the $\bZ$ gauge symmetry
\ie\label{XY-Villain-gaugesym}
\phi \sim \phi + 2\pi k~, \qquad n_\mu \sim n_\mu + \Delta_\mu k~,
\fe
where $k$ is an integer-valued gauge parameter on the sites. We can interpret $n_\mu$ as a $\bZ$ gauge field, which makes $\phi$ compact.

The gauge invariant ``field strength'' of the gauge field $n_\mu$ is
\ie
{\cal N}\equiv\Delta_x n_y-\Delta_y n_x~.
\fe
It can be interpreted as the local vorticity of the configurations.

We are interested is suppressing vortices. One way to do that is to add to the action \eqref{XY-Villain-action} a term like
\ie\label{XY-Villain-vortenergy}
\kappa \sum_\text{plaquette} {\cal N}^2~
\fe
with positive $\kappa$.
For $\kappa\to \infty$ the vortices are completely suppressed \cite{Gross:1990ub}.
Instead of adding this term and taking this limit, we can introduce a Lagrange multiplier $\tilde \phi$ to impose ${\cal N}=0$ as a constraint. The full action now becomes \cite{Sulejmanpasic:2019ytl}\footnote{Related ideas were used in various places, including \cite{Sachdev_2002}.}
\ie\label{XY-modifiedVillain-action}
S=\frac{\beta}{2} \sum_\text{link} (\Delta_\mu \phi - 2\pi n_\mu)^2 + i \sum_\text{plaquette} \tilde \phi {\cal N}~,
\fe
where the Lagrange multiplier $\tilde \phi$ is a real-valued field on the plaquettes (or dual sites).  It has a $\bZ$ gauge symmetry
\ie\label{XY-modifiedVillain-windgaugesym}
\tilde \phi \sim \tilde \phi + 2\pi \tilde k~,
\fe
with $\tilde k$ is an integer-valued gauge parameter on the plaquettes.

Note that the action \eqref{XY-modifiedVillain-action} is not invariant under this gauge symmetry.  However, $e^{-S}$ is gauge invariant.  In fact, even the local quantity $e^{-{\cal L}}$, with $\cal L$ the Lagrangian density, is invariant.

The action \eqref{XY-modifiedVillain-action} is the starting point of our discussion.  We refer to it as the modified Villain action of the XY-model.\footnote{Using common terminology in the condensed matter literature, one could refer to the corresponding theory as noncompact.  However, we emphasize that even though the $\phi$ field in \eqref{XY-Villain-action} and \eqref{XY-modifiedVillain-action} is real-valued, i.e., noncompact, the gauge symmetry \eqref{XY-Villain-gaugesym} effectively compactifies the range of $\phi$.  The effect of the term with $\cal N$ in \eqref{XY-modifiedVillain-action} is to suppress the vortices rather than to de-compactify the target space.  We will discuss it further below.\label{noncompf}}

We can restore the vortices by perturbing the modified Villain action \eqref{XY-modifiedVillain-action} as
\ie\label{XY-modifiedVillain-action-vort}
\frac{\beta}{2} \sum_\text{link} (\Delta_\mu \phi - 2\pi n_\mu)^2 + i \sum_\text{plaquette} \tilde \phi {\cal N} - \lambda \sum_\text{plaquette}\cos(\tilde \phi)~.
\fe
(For simplicity of the presentation, we take $\lambda \ge 0$.) Note that the action is still invariant under the gauge symmetries \eqref{XY-Villain-gaugesym} and \eqref{XY-modifiedVillain-windgaugesym}. Integrating out $\tilde \phi$ gives
\ie
&\frac{\beta}{2} \sum_\text{link} (\Delta_\mu \phi - 2\pi n_\mu)^2 - \sum_\text{plaquette}\log I_{|{\cal N}|}(\lambda)~,
\fe
where $I_k(z)$ is the modified Bessel function of the first kind. Let us compare this action with \eqref{XY-Villain-vortenergy}.  For small $\lambda \ll 1$, we have
\ie
-\log I_k(\lambda) \approx \log \left[ k! \left( \frac{2}{\lambda} \right)^k \right] + O(\lambda^2)~.
\fe
In this case, vortices with ${|{\cal N}|}>1$ are suppressed. For ${|{\cal N}|}=0,1$ we identify
\ie
\kappa \approx \log \frac{2}{\lambda} \gg 1~.
\fe
In the other limit $\lambda \gg 1$, we have
\ie
-\log I_k(\lambda) \sim \frac{1}{2\lambda}k^2 + O(\lambda^{-2})
\fe
where we ignored some $k$-independent terms that depend on $\lambda$. In this case, we can identify
\ie
\kappa \approx \frac{1}{2\lambda} \ll 1~.
\fe

We conclude that the deformation $-\lambda \cos (\tilde \phi)$ is mapped to  $\kappa {\cal N}^2$, and small (large) $\lambda$ corresponds to large (small) $\kappa$.

To summarize, the XY-model is usually studied using the actions \eqref{XY-action} or \eqref{XY-Villain-action}.  We added another coupling to this model \eqref{XY-Villain-vortenergy}.   Equivalently, we can write the model as \eqref{XY-modifiedVillain-action-vort} and then the usually studied model \eqref{XY-Villain-action} is obtained in the limit $\lambda \to \infty$.
On the other hand, when $\lambda=0$, this reduces to our modified Villain action \eqref{XY-modifiedVillain-action} of the XY-model.

Below we will see that the modified Villain action \eqref{XY-modifiedVillain-action}, unlike its other lattice relatives, exhibits many properties similar to its continuum limit, including emergent global symmetries, anomalies, and self-duality.

\subsubsection{Global symmetries}\label{sec:XY-sym}

The three models, \eqref{XY-action}, \eqref{XY-Villain-action}, and \eqref{XY-modifiedVillain-action} have a \emph{momentum symmetry}, which acts as
\ie\label{XY-momsym}
\phi \rightarrow \phi + c^m~,
\fe
where $c^m$ is a real position-independent constant. Due to the zero mode of the gauge symmetry \eqref{XY-Villain-gaugesym}, the $2\pi\bZ$ part of this symmetry is gauged. So the momentum symmetry is $U(1)$ rather than $\bR$.

From \eqref{XY-Villain-action} and \eqref{XY-modifiedVillain-action} we find the Noether current of momentum symmetry\footnote{The factor of $i$ in the Euclidean signature is such that the corresponding charge is real.}
\ie\label{XY-momcur}
J^m_\mu = -i\beta (\Delta_\mu \phi - 2\pi n_\mu)~,
\fe
which is conserved because of the equation of motion of $\phi$.  The momentum charge is\footnote{Here, $\epsilon_{xy} = - \epsilon_{yx} = 1$ and $\epsilon_{xx} = \epsilon_{yy}=0$.}
\ie\label{Qmchar}
Q^m(\tilde{\mathcal C}) = \sum_{\text{dual link}\in \tilde{\mathcal C}} \epsilon_{\mu\nu}J^m_\nu~,
\fe
where $\tilde{\mathcal C}$ is a curve along the dual links of the lattice. The dependence of $Q^m$ on $\tilde{\mathcal C}$ is topological.  The local operator $e^{i\phi}$ is charged under this symmetry.

The modified Villain action \eqref{XY-modifiedVillain-action} (but not \eqref{XY-action} or \eqref{XY-Villain-action}) also has a \emph{winding symmetry}, which acts as
\ie\label{XY-modifiedVillain-windsym}
\tilde \phi \rightarrow \tilde \phi + c^w~,
\fe
where $c^w$ is a real constant. Due to the zero mode of the gauge symmetry \eqref{XY-modifiedVillain-windgaugesym}, the $2\pi \bZ$ part of this symmetry is gauged. So the winding symmetry is also $U(1)$.

The Noether current of the winding symmetry is\footnote{From the action \eqref{XY-modifiedVillain-action}, the Noether current appears to be $J^w_\mu=-\epsilon_{\mu\nu}n_\nu$, but it is not gauge invariant. Therefore, we added to it an improvement term to construct a gauge invariant current.}
\ie\label{XY-modifiedVillain-windcur}
J^w_\mu = \frac{\epsilon_{\mu\nu}}{2\pi} (\Delta_\nu \phi - 2\pi n_\nu)~,
\fe
which is conserved because of the equation of motion of $\tilde \phi$. It is crucial that $n_\mu$ is flat, i.e., ${\cal N}=0$ and vortices are suppressed, for the Noether current to be conserved. The winding charge is
\ie\label{XY-modifiedVillain-windcharge}
Q^w(\mathcal C) = \sum_{\text{link}\in \mathcal C} \epsilon_{\mu\nu}J^w_\nu = -\sum_{\text{link}\in \mathcal C} n_\mu~,
\fe
where $\mathcal C$ is a curve along the links of the lattice. The last equation follows from the single-valuedness of $\phi$. Hence, we can interpret $Q^w(\mathcal C)$ as the gauge invariant Wilson line of the $\bZ$ gauge field $n_\mu$. It is topological due to the flatness condition of $n_\mu$.   Finally, the local operator $e^{i\tilde \phi}$ is charged under this symmetry.

Both the momentum symmetry \eqref{XY-momsym} and the winding symmetry \eqref{XY-modifiedVillain-windsym} act locally on the fields and they both leave the action \eqref{XY-modifiedVillain-action} invariant.  However, the Lagrangian density in \eqref{XY-modifiedVillain-action} is invariant under the momentum symmetry, but not under the winding symmetry.  This fact makes it possible for these symmetries  to have a mixed 't Hooft anomaly, even though the two symmetries act locally (``on-site'').

Using ``summing by parts'', we can write \eqref{XY-modifiedVillain-action} as
\ie\label{summbp}
\frac{\beta}{2} \sum_\text{link} (\Delta_\mu \phi - 2\pi n_\mu)^2 + i \sum_\text{plaquette} (n_x\Delta_y \tilde \phi - n_y \Delta_x \tilde \phi)~.
\fe
In this form both the momentum symmetry \eqref{XY-momsym} and the winding symmetry \eqref{XY-modifiedVillain-windsym} act locally and leave the Lagrangian density invariant.  How is this compatible with the anomaly?  The point is that unlike \eqref{XY-modifiedVillain-action}, the Lagrangian density in \eqref{summbp} is not gauge invariant.  As is common with anomalies, we can move the problem around, but we cannot completely avoid it.

One way to see this anomaly is by trying to couple the action \eqref{XY-modifiedVillain-action} to background gauge fields for the momentum and winding symmetries $(A_\mu;N)$ and $(\tilde A_\mu;\tilde N)$.
Here $A_\mu,\tilde A_\mu$ are real-valued and $N,\tilde N$ are integer-valued.
The action is
\ie\label{anomaly}
&\frac{\beta}{2} \sum_\text{link} (\Delta_\mu \phi - A_\mu - 2\pi n_\mu)^2 + i \sum_\text{plaquette} \tilde \phi (\Delta_x n_y - \Delta_y n_x + N)
\\
& - \frac{i}{2\pi}\sum_\text{link} \epsilon_{\mu\nu} \tilde A_\mu (\Delta_\nu \phi - A_\nu - 2\pi n_\nu) + i\sum_\text{site} \tilde N \phi~,
\fe
with the gauge symmetry
\ie
&\phi \sim \phi + \alpha + 2\pi k~, && \tilde \phi \sim \tilde \phi + \tilde \alpha + 2\pi \tilde k~,
\\
&A_\mu \sim A_\mu + \Delta_\mu \alpha + 2\pi K_\mu~, && \tilde A_\mu \sim \tilde A_\mu + \Delta_\mu \tilde \alpha + 2\pi \tilde K_\mu~,
\\
&n_\mu \sim n_\mu + \Delta_\mu k - K_\mu~, && \tilde N \sim \tilde N + \Delta_x \tilde K_y - \Delta_y \tilde K_x~,
\\
&N \sim N + \Delta_x K_y - \Delta_y K_x~.\qquad
\fe
Here, $K_\mu,\tilde K_\mu$ are integers, and $\alpha,\tilde \alpha$ are real.  They are the gauge parameters of the background gauge fields $(A_\mu;N)$ and $(\tilde A_\mu;\tilde N)$. The variation of the action under this gauge transformation is
\ie\label{anomalyuouo}
- \frac{i}{2\pi}\sum_\text{plaquette} \tilde \alpha (\Delta_x A_y - \Delta_y A_x - 2\pi N) + i\sum_\text{plaquette} (\tilde K_x A_y - \tilde K_y A_x)  + i\sum_\text{site} (\tilde N + \Delta_x \tilde K_y - \Delta_y \tilde K_x) \alpha~.
\fe
It signals an anomaly because it cannot be cancelled by adding any 1+1d local counterterms.  This expression of the anomaly is the lattice version of the familiar continuum expression $- \frac{i}{2\pi}\int dxdy~ \tilde \alpha (\partial_x A_y - \partial_y A_x)$.

As a special case of this anomaly, consider the $\mathbb{Z}_N$ subgroup of the $U(1)\times U(1)$ symmetry, which is generated by $\phi\to \phi+2\pi/N,\ \tilde\phi\to\tilde\phi +2\pi/N$.  The anomaly in this symmetry is visible in \eqref{anomalyuouo}.  It agrees with the general classification of $\mathbb{Z}_N$ anomalies in 1+1d bosonic systems by $H^3(\mathbb{Z}_N,U(1))=\mathbb{Z}_N$.

\subsubsection{T-Duality}\label{XYTduality}

Here we will demonstrate the self-duality of the modifield Villain lattice model \eqref{XY-modifiedVillain-action}.
We start with the presentation \eqref{summbp}.  Using the Poisson resummation formula \eqref{Possonresummationi} for $n_\mu$ and ignoring the overall factor, we can dualize the above action to
\ie\label{dualmoVXY}
&\frac{1}{2(2\pi)^2 \beta} \sum_\text{dual link} (\Delta_\mu \tilde \phi - 2\pi \tilde n_\mu)^2 + i \sum_\text{site} \phi \tilde {\cal N}~,\\
&\tilde {\cal N}\equiv \Delta_x \tilde n_y - \Delta_y \tilde n_x~,
\fe
where $\tilde n_\mu$ is an integer-valued field on the dual links. The gauge symmetry of the original theory acts as
\ie
\tilde \phi \sim \tilde \phi + 2\pi \tilde k~,\qquad \tilde n_\mu \sim \tilde n_\mu + \Delta_\mu \tilde k,\qquad \phi\sim \phi+2\pi k~.
\fe
$\tilde n_\mu$ can be interpreted as the $\bZ$ gauge field associated with the gauge symmetry of $\tilde \phi$ and $\tilde {\cal N}$ is its field strength. Furthermore, we can interpret $\phi$ as a Lagrange multiplier imposing  $\tilde {\cal N}=0$ as a constraint.

We conclude that  the modified Villain action \eqref{XY-modifiedVillain-action} is a self-dual lattice model with $\beta \leftrightarrow \frac{1}{(2\pi)^2 \beta}$. Moreover, the momentum and winding currents, \eqref{XY-momcur} and \eqref{XY-modifiedVillain-windcur}, in the dual picture are
\ie\label{currentsdua}
J^m_\mu = \frac{\epsilon_{\mu\nu}}{2\pi} (\Delta_\nu \tilde \phi - 2\pi \tilde n_\nu)~,\qquad J^w_\mu = -\frac{i}{(2\pi)^2\beta} (\Delta_\mu \tilde \phi - 2\pi \tilde n_\mu)~.
\fe

We emphasize that the lattice model \eqref{XY-modifiedVillain-action} is exactly self-dual, rather than being only IR-self-dual.  It has exact T-duality.

We can easily relate this discussion to the classical analysis of \cite{PhysRevB.16.1217,PhysRevB.17.1340}.  By adding the term $-\lambda \cos (\tilde\phi) $ to the Lagrangian and taking $\lambda\to \infty$, the field $\tilde \phi$ is frozen at zero and we end up with Villain action \eqref{XY-Villain-action}.  Repeating this in the dual action \eqref{dualmoVXY}, we find
\ie
&\frac{1}{2 \beta} \sum_\text{dual link} \tilde n_\mu^2 + i \sum_\text{site} \phi \tilde {\cal N}~,\\
&\tilde {\cal N}\equiv \Delta_x \tilde n_y - \Delta_y \tilde n_x~.
\fe
Locally, the Lagrange multiplier $\phi$ determines $\tilde n_\mu = \Delta_\mu q$ with an integer $q$.\footnote{More precisely, $\tilde {\cal N}=0$ can be solved in terms of an integer-valued field $q$, but $q$ does not have to be periodic (i.e., single-vlaued on the torus).  Its lack of periodicity is characterized by two integers, which are the Wilson lines of $\tilde n$ around two cycles of the torus. This Wilson line is the momentum charge \eqref{Qmchar} constructed out of the momentum current \eqref{currentsdua} and it is nontrivial only when $q$ is not periodic.}  We end up with
\ie
\frac{1}{2 \beta} \sum_\text{dual link} (\Delta_\mu q)^2 ~,
\fe
which is the dual theory of \cite{PhysRevB.16.1217,PhysRevB.17.1340}.

\subsubsection{Gauge-fixing and the continuum limit}

In the following we will pick a convenient gauge where most of the integer fields are set to zero.
Following the discussion around \eqref{gaugecom}, we integrate out $\tilde \phi$, which imposes the flatness condition on $n_\mu$.  Then, we gauge fix $n_\mu(\hat x,\hat y)=0$ at all links, except $n_x(L^x-1,\hat y)$ and $n_y(\hat x,L^y-1)$ (recall, $\hat x^\mu \sim \hat x^\mu + L^\mu$).  The remaining information in the gauge fields $n_\mu$ is in the two integers $n_x(L^x-1,\hat y) \equiv \bar n_x$ and $n_y(\hat x,L^y-1) \equiv \bar n_y$, i.e., in the holonomies of $n_\mu$ around the $x$ and $y$ cycles.
The residual gauge symmetry is
\ie
\phi \sim \phi + 2\pi \bZ~.
\fe

Let us define a new field $\bar \phi$ such that
\ie
\bar\phi(0,0) = \phi(0,0)~,\qquad \Delta_\mu \bar \phi = \Delta_\mu \phi - 2\pi n_\mu~.
\fe
In the gauge above, where in most of the links $n_\mu=0$, in most of the sites $\bar\phi=\phi$.
Then the action in terms of $\bar \phi$ is
\ie\label{XY-modifiedVillain-gaugefix-action}
\frac{\beta}{2} \sum_\text{link} (\Delta_\mu \bar \phi)^2~.
\fe
Although $\phi$ and $n_\mu$ are single-valued fields, $\bar \phi$ can wind around nontrivial cycles:
\ie\label{twistbe}
&\bar \phi(\hat x + L^x,\hat y) = \bar \phi(\hat x,\hat y) - 2\pi \bar n_x~,
\\
&\bar \phi(\hat x,\hat y + L^y) = \bar \phi(\hat x,\hat y) - 2\pi \bar n_y~.
\fe
So, in the path integral, we should sum over nontrivial winding sectors of $\bar \phi$.\footnote{Note that the variables $\bar\phi$ are noncompact and we can rescale them to make the action \eqref{XY-modifiedVillain-gaugefix-action} independent of $\beta$. Then, the compactness and the $\beta$ dependence enter only through the twisted boundary conditions \eqref{twistbe}.  One might say that therefore, the local dynamics is independent of $\beta$ and the model is the same as that of a noncompact scalar.  This is the rationale behind the terminology mentioned in footnote \eqref{noncompf}.  This reasoning is valid when we consider the model with fixed twisted boundary conditions like \eqref{twistbe}.   However, in our case, we sum over this twist.  And this affects the set of local operators in the theory.  In particular, as in \eqref{opspec}, their dimensions depend on the value of $\beta=R^2 /\pi$.\label{compcont}}

In the continuum limit $a\rightarrow 0$ such that $\ell^\mu \equiv aL^\mu$ is fixed, the action \eqref{XY-modifiedVillain-gaugefix-action} becomes
\ie
\frac{\beta}{2} \int dxdy ~ (\partial_\mu \phi)^2~,
\fe
where we dropped the bar on $\phi$. This is the action of the 2d compact boson.  Locally, this is the same as a theory of a noncompact scalar $\phi$.  However, here we sum over twisted boundary conditions and that makes the $\phi$ field compact.  See the related discussion in footnote \ref{compcont}.

\begin{figure}[t]
\centering
\includegraphics[scale=0.5]{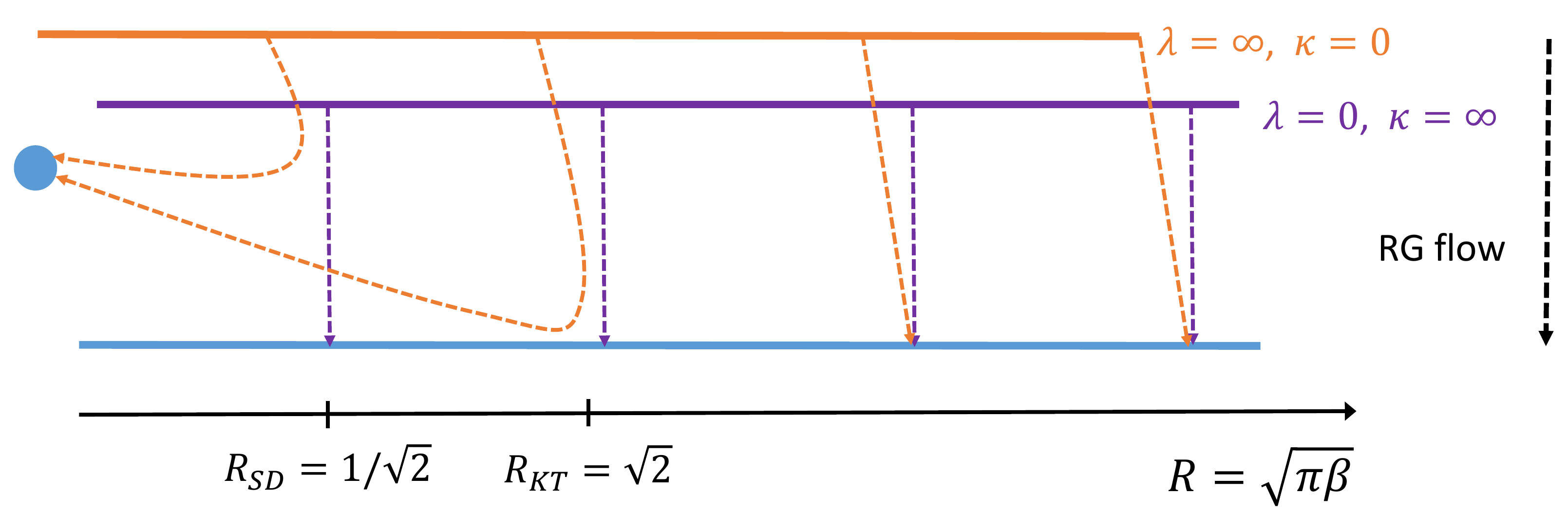}
\caption{The space of coupling constants of the 2d Euclidean XY-model.  The orange line corresponds to the theories based on \eqref{XY-action} or \eqref{XY-Villain-action}, while the purple line corresponds to the modified theory \eqref{XY-modifiedVillain-action}.  Each of them depends on the parameter $R=\sqrt{\pi\beta}$.  The parameter $\lambda$ (equivalently, $\kappa$) interpolates between these two lines.  The theories of the purple line \eqref{XY-modifiedVillain-action} are special because they have a global $U(1)$ winding symmetry and they enjoy a $R\to {1\over 2R}$ duality with selfduality at $R={1\over \sqrt 2}$.  The dashed lines represent the renormalization group flow, or equivalently the continuum limit.  The theories of the purple line
flow to the $c=1$ compact-boson conformal field theories, which are represented by the blue line.  The theories of the orange line \eqref{XY-action} or \eqref{XY-Villain-action} also flow to this conformal theory, provided $R\ge R_{KT}=\sqrt 2$ (equivalently, $\beta\ge {2\over \pi}$).  For $R< R_{KT}=\sqrt 2$ (equivalently, $\beta< {2\over \pi}$), the theories of the orange line flow to a gapped phase, which is represented by the blue region at the left.  The more generic theories with nonzero but finite $\lambda$ (and $\kappa$) behave like the theories of the orange line.}
\label{fig:XY-diagram}
\end{figure}

\subsubsection{Kosterlitz-Thouless transition}

In order to compare with the standard conformal field theory literature (e.g., \cite{Ginsparg:1988ui,DiFrancesco:1997nk}), we define the radius $R$ of the compact boson as $R=\sqrt{\pi\beta}$. The theory at radius $R$ has momentum and winding operators with dimensions
\ie\label{opspec}
(h,\bar h)=\left(\frac{1}{2}\left(\frac{n_m}{2R}+n_wR\right)^2,\frac{1}{2}\left(\frac{n_m}{2R}-n_wR\right)^2\right)~,
\fe
where $n_m,n_w$ are the momentum and winding charges of the operator. These operators correspond to the lattice operators $e^{i(n_m\phi+n_w\tilde \phi)}$.  T-duality exchanges the theories at radius $R$ and $\frac{1}{2R}$. At the radius $R=\frac{1}{\sqrt{2}}$, the theory is self-dual.  See Figure \ref{fig:XY-diagram}.

Unlike the modified Villain model \eqref{XY-modifiedVillain-action}, the original XY-model \eqref{XY-action} and its Villain counterpart \eqref{XY-Villain-action} have only the momentum symmetry, but no winding symmetry. It could still happen that their long-distance theory has such an emergent winding symmetry. This happens when the winding number violating operators are irrelevant (or exactly marginal) in the IR theory.  This is the case for $R\ge R_{KT}=\sqrt 2$, or equivalently $\beta\ge\beta_{KT}= \frac{2}{\pi}$, where the subscript $KT$ stands for Kosterlitz-Thouless.  However, for smaller values of $R$ and $\beta$ the winding operators are relevant and the lattice models undergo the Kosterlitz-Thouless transition to a gapped phase. See Figure \ref{fig:XY-diagram}.

Finally,  this reasoning implies that the qualitative behavior of the flow for finite nonzero $\lambda$ is the same as the flow for infinite $\lambda$ in Figure \ref{fig:XY-diagram}.  Only for $\lambda=0$ is the flow different (as the purple line in Figure \ref{fig:XY-diagram}).  Also, it is straightforward to replace the deformation $\cos (\tilde \phi)$ by $\cos (W\tilde \phi)$ for generic integer $W$.  This breaks the $U(1)$ winding symmetry to $\bZ_W$.  Then the flow is as from the orange curve in Figure \ref{fig:XY-diagram}, except that the Kosterlitz-Thouless point moves to $R={\sqrt 2\over W}$.

\subsection{2d Euclidean $\mathbb{Z}_N$ clock model}\label{sec:ZNclock}

\subsubsection{Lattice models}

The $\mathbb{Z}_N$ clock model \cite{PhysRevB.16.1217,PhysRevD.19.3698,KADANOFF197939,Cardy_1980,Alcaraz:1980sa,FRADKIN19801} can be obtained by restricting the phase variables $e^{i\phi}$ in the XY-model \eqref{XY-action} to $\mathbb{Z}_N$ variables $e^{2\pi i m/N}$.  More generally, this model has $\lfloor N/2\rfloor$ nearest-neighbor couplings
\ie\label{ZNclock}
 \sum_{M=1}^{\lfloor N/2\rfloor } J_M \sum_{\text{link}} \left[1-\cos\left(\frac{2\pi M}{N}\Delta_\mu m\right)\right]~.
\fe
where $\lfloor N/2\rfloor $ is the integer part of $N/2$.
A particular one-dimensional locus in the parameter space of $\{J_M\}$ is given by the Villain action:
\ie\label{eq:ZNVillain}
\frac{\beta}{2}\left(\frac{2\pi}{N}\right)^2\sum_{\text{link}}\left(\Delta_\mu m-Nn_\mu\right)^2~.
\fe
The integer fields $m,n_\mu$ are subject to a gauge symmetry with integer gauge parameter $k$
\ie
&m\sim m+Nk~,
\\
&n_\mu\sim n_\mu+\Delta_\mu k~.
\fe

This model \eqref{eq:ZNVillain} can be embedded in the XY-model of Appendix \ref{2dXY}.  In general, we can deform the action \eqref{XY-modifiedVillain-action} to
\ie\label{ZNclocklambdat}
 \frac{\beta}{2} \sum_\text{link} (\Delta_\mu \phi - 2\pi n_\mu)^2 + i \sum_\text{plaquette} \tilde \phi {\cal N} - \lambda\sum_\text{plaquette}\cos(W\tilde\phi)-\tilde\lambda\sum_\text{site}\cos(N\phi)~,
\fe
with integer $N$ and $W$.  The term with $\tilde \lambda$ breaks the $U(1)$ momentum global symmetry to $\mathbb{Z}_N$, which is generated by $\phi \to \phi + {2\pi\over N}$.  Similarly, the term with $ \lambda$
breaks the $U(1)$ winding global symmetry to $\mathbb{Z}_W$.

The most commonly analyzed case is with $W=1$ and $\tilde \lambda, \lambda \to \infty$.  Then, $\tilde\phi$ is constrained to vanish and therefore the vortices are not suppressed.  Similarly, $\phi $ is constrained to have the values $\phi ={2\pi m \over N}$, thus leading to \eqref{eq:ZNVillain}.

\subsubsection{Kramers-Wannier duality}\label{KWduality}

It is straightforward to repeat the analysis in Appendix \ref{XYTduality} and to dualize \eqref{ZNclocklambdat} to
\ie\label{dualZN}
&\frac{1}{2(2\pi)^2 \beta} \sum_\text{dual link} (\Delta_\mu \tilde \phi - 2\pi \tilde n_\mu)^2 + i \sum_\text{site} \phi \tilde {\cal N}- \lambda\sum_\text{plaquette}\cos(W\tilde\phi)-\tilde\lambda\sum_\text{site}\cos(N\phi)~,\\
&\tilde {\cal N}\equiv \Delta_x \tilde n_y - \Delta_y \tilde n_x~,
\fe
where $\tilde n_\mu$ is an integer-valued field on the dual links. The gauge symmetry of the theory is
\ie
\tilde \phi \sim \tilde \phi + 2\pi \tilde k~,\qquad \tilde n_\mu \sim \tilde n_\mu + \Delta_\mu \tilde k,\qquad \phi\sim \phi+2\pi k~.
\fe

We conclude that the action \eqref{ZNclocklambdat} is dual to a similar system with  $\beta \leftrightarrow \frac{1}{(2\pi)^2 \beta}$ and $N\leftrightarrow W$.

In the special case with $W=1$ and $\tilde \lambda, \lambda \to \infty$, \eqref{ZNclocklambdat} is dualized to
\ie\label{dualZNs}
\frac{1}{2 \beta} \sum_\text{dual link} \tilde n_\mu^2 + {2\pi i \over N}\sum_\text{site} m (\Delta_x \tilde n_y - \Delta_y \tilde n_x)
\fe
with the gauge symmetry
\ie\label{mgauge}
m\sim m+N k~
\fe
with integer $k$.
We can find it either by substituting $\phi ={2\pi m \over N}$, $\tilde \phi=0$ in \eqref{dualZN}, or by directly dualizing \eqref{eq:ZNVillain}.

We see that unlike the modified Villain action for the XY-model \eqref{XY-modifiedVillain-action}, this theory is not selfdual.  Comparing with the general case \eqref{dualZN}, this follows from the fact that now $W=1$ and the duality there exchanges $W\leftrightarrow N$.

How is this consistent with the known Kramers-Wannier duality of this theory \cite{PhysRevB.16.1217,PhysRevD.19.3698,KADANOFF197939,Cardy_1980,Alcaraz:1980sa,FRADKIN19801}?

In order to answer this question we first add integer-valued fields $\tilde m$ and $\hat n_\mu$ to the action \eqref{dualZNs}
\ie\label{dualZNg}
\frac{1}{2 \beta} \sum_\text{dual link} (\Delta_\mu \tilde m - N \hat n_\mu -\tilde n_\mu)^2 + {2\pi i \over N}\sum_\text{site} m (\Delta_x \tilde n_y - \Delta_y \tilde n_x)~.
\fe
In addition to the gauge symmetry \eqref{mgauge}, this action has the gauge symmetry
\ie\label{newgas}
&\tilde m\sim\tilde m+\tilde k~,\\
&\hat n_\mu\sim\hat n_\mu-\hat q_\mu~,\\
&\tilde n_\mu\sim\tilde n_\mu+\Delta_\mu\tilde k+N\hat q_\mu~.
\fe
Here $\tilde k$ is an integer zero-form gauge parameter and $\hat q_\mu$ is an integer one-form gauge parameter. This new action \eqref{dualZNg} is equivalent to \eqref{dualZNs}, as can be seen by completely gauge fixing \eqref{newgas} by setting  $\tilde m =\hat n_\mu=0$.

Now, we can interpret \eqref{dualZNg} as follows.  Locally, the Lagrange multiplier $m$ sets $\tilde n_\mu$ to a pure gauge and we can set it to zero. Then, \eqref{dualZNg} is the same as the Villain form of the $\bZ_N$ action \eqref{eq:ZNVillain} with the replacement $\beta\leftrightarrow\frac{N^2}{4\pi^2\beta}$.  This shows that locally, the $\bZ_N$ clock-model has Kramers-Wannier duality.

However, globally, the Lagrange multiplier $m$ in \eqref{dualZNg} does not set $\tilde n_\mu$ to a pure gauge and it allows configurations with nontrivial holonomies $\sum_{\text{links}} n_\mu$ around closed cycles.  In other words, \eqref{dualZNg} is not a $\bZ_N$ clock-model but a $\bZ_N$ clock-model coupled to a topological lattice $\bZ_N$ gauge theory \cite{Dijkgraaf:1989pz,Kapustin:2014gua}.
The latter is described by the second term in \eqref{dualZNg} and will be further discussed  in Appendix \ref{ZNgauget}.

We conclude that the T-duality of the underlying XY-model \eqref{XY-modifiedVillain-action} leads to the Kramers-Wannier duality of the clock-model \eqref{eq:ZNVillain}.  In fact, while the T-duality is correct both locally and globally, the Kramers-Wannier duality of the clock-model is valid also globally only when a lattice topological theory is included in one side of the duality.

\subsubsection{Long-distance limit}

Here, we study the long-distance limit of the theory based on \eqref{ZNclocklambdat}.

As in the discussion around Figure \ref{fig:XY-diagram} we start with the theory with $\lambda=\tilde \lambda =0$.  It flows to the compact-boson theory, which is represented by the blue line in Figure \ref{fig:XY-diagram}.  Then, for small enough $\lambda$ and $\tilde \lambda$ we can perturb this conformal theory by these two perturbations.
The momentum breaking operator $\cos(N\phi)$ is irrelevant for $R<{N\over \sqrt 8}$ and the winding breaking operator is irrelevant for $R>{\sqrt 2\over W}$.  Therefore, for $NW\ge 4$ there are values of $R$, or equivalently of $\beta={R^2\over \pi}$, such that the compact-boson conformal field theory is robust under deformations with small $\lambda$ and $\tilde \lambda$.  This happens for
\ie
&{\sqrt 2\over W}\le R\le {N\over \sqrt 8}\\
&{2\over \pi W^2} \le \beta \le { N^2 \over 8\pi}
\fe
and then the long distance theory is gapless.  Note that this is consistent with the duality  $\beta \leftrightarrow \frac{1}{(2\pi)^2 \beta}$, which is accompanied with $N\leftrightarrow W$.

In the most studied case of $W=1$, the long distance theory of \eqref{ZNclocklambdat} is given by the compact scalar CFT for $N\ge 4$.  For $N=4$ and $R=\sqrt 2 $ it is the CFT of the Kosterlitz-Thouless point.  And for $N\ge 5$  and
\ie
&{\sqrt 2}\le R\le {N\over \sqrt 8}\\
&{2\over \pi } \le \beta \le { N^2 \over 8\pi}
\fe
it is the line of a CFT with this value of $R$.  For other values of $R$ the theory is gapped.  Note, as a check that this is consistent with the $R\leftrightarrow {N\over 2R}$ duality of the local dynamics, which we discussed in Appendix \ref{KWduality}.

For $N=2$ and $N=3$ the duality determines that the theory has two gapped phases separated by a CFT at $R=1$ and $R=\sqrt{3\over 2}$, respectively.
However, these CFTs are not the CFT of the compact boson, but are of the Ising and 3-states Potts model.

We should emphasize that this discussion of the clock-model is specific to the action \eqref{ZNclocklambdat}.  For other actions, the gapless phase could be different or even absent.  See the discussion in \cite{Cardy_1980,Alcaraz:1980sa,Fateev:1985mm}.

\section{Modified Villain formulation of $p$-form lattice gauge theory in diverse dimensions}\label{pformgau}

In this appendix, we will study $p$-form gauge theories on a $d$-dimensional Euclidean space for $p\leq d-1$ (see \cite{Savit:1979ny} for a review on these models).
The modified Villain version of the $p$-form $U(1)$ gauge theory in general dimensions has been analyzed in \cite{Sulejmanpasic:2019ytl}.
The models in Appendix \ref{sec:VillainQM}, correspond to $d=1$ and $p=0$ and perhaps do not deserve to be called gauge theories.  The models in Appendix \ref{Villainlattia}, correspond to $d=2$ and $p=0$.

As above, the lattice spacing is $a$, and there are $L^\mu$ sites in the $\mu$ direction. Throughout this discussion, $A^{(p)}$ denotes a $p$-form field placed on the $p$-cells of the lattice, and $\tilde B^{(d-p)}$ denotes a $(d-p)$-form field placed on the dual $(d-p)$-cells.

\subsection{$U(1)$ gauge theory}\label{AppeUone}
Let us place $U(1)$ variables $e^{ia^{(p)}}$ on $p$-cells of the $d$-dimensional Euclidean lattice. The standard action of this gauge field is
\ie\label{U1-pform-action}
\beta \sum_{(p+1)\text{-cell}} [1-\cos(\Delta a^{(p)})]~,
\fe
where $\Delta a^{(p)}$ is a $(p+1)$-form given by the oriented sum of $a^{(p)}$ along the $p$-cells in the boundary of the $(p+1)$-cell, and $a^{(p)}$ is circle-valued with gauge symmetry
\ie
e^{ia^{(p)} }\sim e^{ia^{(p)} + i\Delta \alpha^{(p-1)}} ~,
\fe
where $\alpha^{(p-1)}$ is  circle-valued.    At large $\beta$, the action can be approximated by the Villain action \cite{BANKS1977493,PhysRevD.16.3040,PhysRevLett.39.55}
\ie\label{U1-pform-Vill-action}
\frac \beta 2 \sum_{(p+1)\text{-cell}} (\Delta a^{(p)} - 2\pi n^{(p+1)})^2~,
\fe
where now $a^{(p)}$ is real and $n^{(p+1)}$ is integer-valued. We can interpret $n^{(p+1)}$ as the $\mathbb Z$ gauge field that makes $a^{(p)}$ compact because of the gauge symmetry
\ie\label{U1-pform-gaugesym}
&a^{(p)} \sim a^{(p)} + \Delta \alpha^{(p-1)} + 2\pi k^{(p)}~,
\\
&n^{(p+1)} \sim  n^{(p+1)} + \Delta k^{(p)}~.
\fe

For $p\leq d-2$, nonzero $\Delta n^{(p+1)}$ corresponds to monopoles or vortices. They can be suppressed by modifying \eqref{U1-pform-Vill-action} to \cite{Sulejmanpasic:2019ytl}
\ie\label{U1-pform-modVill-action}
\frac \beta 2 \sum_{(p+1)\text{-cell}} (\Delta a^{(p)} - 2\pi n^{(p+1)})^2 + i \sum_{(p+2)\text{-cell}} \tilde a^{(d-p-2)} \Delta n^{(p+1)}~,
\fe
where $\tilde a^{(d-p-2)}$ is a real-valued $(d-p-2)$-form field, which acts as a Lagrange multiplier imposing the flatness constraint of $n^{(p+1)}$.
We will refer to \eqref{U1-pform-modVill-action} as the modified Villain action of the $U(1)$ $p$-form gauge theory.
In addition to \eqref{U1-pform-gaugesym}, this theory also has a gauge symmetry
\ie\label{U1-pform-modVill-maggaugesym}
\tilde a^{(d-p-2)} \sim \tilde a^{(d-p-2)} + \Delta \tilde \alpha^{(d-p-3)} + 2\pi \tilde k^{(d-p-2)}~,
\fe
where $\tilde \alpha^{(d-p-3)}$ is real-valued, and $\tilde k^{(d-p-2)}$ is integer-valued.

For $p=d-1$ we cannot write \eqref{U1-pform-modVill-action}.  Instead, in this case we can add another term\footnote{See, for example, \cite{Gattringer:2018dlw,Sulejmanpasic:2019ytl,Anosova:2019quw,Sulejmanpasic:2020lyq,Sulejmanpasic:2020ubo} for discussions on the $\theta$-angle in the Villain version of the lattice $U(1)$ gauge theory. }
\ie\label{U1-pform-modVill-actiont}
\frac \beta 2 \sum_{d\text{-cell}} (\Delta a^{(d-1)} - 2\pi n^{(d)})^2 + i \theta \sum_{d\text{-cell}} n^{(d)}~.
\fe
This is a $U(1)$ gauge theory of a $(d-1)$-form gauge field with a $\theta$-parameter. (Compare with \eqref{particlerL} and \eqref{altLaqm}, which corresponds to $p=0$ and $d=1$.) Note that this is a lattice version of the gauge theory with $\theta$.  Unlike the continuum presentation, here, the $\theta$-term is associated with the integer-valued field.  The topological charge $\sum_{d\text{-cell}} n^{(d)}$ is manifestly quantized and therefore  $\theta \sim \theta+2\pi$.

\subsubsection{Duality}
Using the Poisson resummation formula \eqref{Possonresummationi}, we can dualize the modified Villain action \eqref{U1-pform-modVill-action} of a $p$-form gauge theory to the modified Villain action of a $(d-p-2)$-form gauge theory
\ie
\frac{1}{2(2\pi)^2\beta} \sum_{(p+1)\text{-cell}} (\Delta \tilde a^{(d-p-2)} - 2\pi \tilde n^{(d-p-1)})^2 + i(-1)^{d-p} \sum_{(p+1)\text{-cell}} \tilde n^{(d-p-1)} \Delta a^{(p)}~,
\fe
where $\tilde n^{(d-p-1)}$ is integer-valued. We can interpret $\tilde n^{(d-p-1)}$ as a $\mathbb Z$ gauge field that makes $\tilde a^{(d-p-2)}$ compact because of the gauge symmetry
\ie
&\tilde a^{(d-p-2)} \sim \tilde a^{(d-p-2)} + \Delta \tilde \alpha^{(d-p-3)} + 2\pi \tilde k^{(d-p-2)}~,
\\
&\tilde n^{(d-p-1)} \sim \tilde n^{(d-p-1)} + \Delta \tilde k^{(d-p-2)}~.
\fe
The field $a^{(p)}$ is a Lagrange multiplier that imposes the flatness constraint of $\tilde n^{(d-p-1)}$. When $d$ is even, and $p = \frac{d-2}2$, the model \eqref{U1-pform-modVill-action} is self-dual with $\beta \leftrightarrow \frac{1}{(2\pi)^2\beta}$.

\subsubsection{Global symmetries}

In all the three models, \eqref{U1-pform-action}, \eqref{U1-pform-Vill-action}, and \eqref{U1-pform-modVill-action}, there is a $p$-form \emph{electric symmetry} \cite{Gaiotto:2014kfa}, which acts on the fields as
\ie
a^{(p)} \rightarrow a^{(p)} + \lambda^{(p)}~,
\fe
where $\lambda^{(p)}$ is a real-valued, flat $p$-form field. Due to the gauge symmetry \eqref{U1-pform-gaugesym}, the electric symmetry is $U(1)$ rather than $\bR$. In \eqref{U1-pform-Vill-action} and \eqref{U1-pform-modVill-action}, the Noether current of electric symmetry is\footnote{The Hodge dual $\star A^{(p)}$ is a $(d-p)$-form field on the dual $(d-p)$-cells of the lattice.}
\ie
J_e^{(p+1)} = i \beta (\Delta a^{(p)} - 2\pi n^{(p+1)}) = \frac{(-1)^{d-p}}{2\pi} \star (\Delta \tilde a^{(d-p-2)} - 2\pi \tilde n^{(d-p-1)})~,
\fe
which is conserved because of the equation of motion of $a^{(p)}$. The electric charge is
\ie
Q_e(\tilde{\mathcal M}^{(d-p-1)}) = \sum_{\text{dual }(d-p-1)\text{-cell}\in \tilde{\mathcal M}^{(d-p-1)}} \star J_e^{(p+1)}~,
\fe
where $\tilde{\mathcal M}^{(d-p-1)}$ is a codimension-$(p+1)$ submanifold along the dual $(d-p-1)$-cells of the lattice. The electrically charged objects are the Wilson observables
\ie
W_e(\mathcal M^{(p)}) = \exp\left[ i \sum_{p\text{-cell}\in \mathcal M^{(p)}} a^{(p)} \right]~,
\fe
where $\mathcal M^{(p)}$ is a dimension-$p$ submanifold along the $p$-cells of the lattice.

The theory \eqref{U1-pform-modVill-action} (but not \eqref{U1-pform-action} or \eqref{U1-pform-Vill-action}) also has a $(d-p-2)$-form \emph{magnetic symmetry} \cite{Gaiotto:2014kfa}, which acts on the fields as
\ie
\tilde a^{(d-p-2)} \rightarrow \tilde a^{(d-p-2)} + \tilde \lambda^{(d-p-2)}~,
\fe
where $\tilde \lambda^{(d-p-2)}$ is a real-valued, flat $(d-p-2)$-form. Due to the gauge symmetry \eqref{U1-pform-modVill-maggaugesym}, the magnetic symmetry is $U(1)$. The Noether current of magnetic symmetry is\footnote{Recall that $\star \star A^{(p)} = (-1)^{p(d-p)} A^{(p)}$.}
\ie
J_m^{(d-p-1)} = - \frac{i}{(2\pi)^2\beta} \star \star(\Delta \tilde a^{(d-p-2)} - 2\pi \tilde n^{(d-p-1)}) = \frac{(-1)^{d-p}}{2\pi} \star (\Delta a^{(p)} - 2\pi n^{(p+1)})~,
\fe
which is conserved because of the equation of motion of $\tilde a^{(d-p-2)}$. The magnetic charge is
\ie
Q_m(\mathcal M^{(p+1)}) = \sum_{(p+1)\text{-cell}\in \mathcal M^{(p+1)}} \star J_m^{(d-p-1)}~,
\fe
where $\mathcal M^{(p+1)}$ is a dimension-$(p+1)$ submanifold along the $(p+1)$-cells of the lattice. The magnetically charged objects are the 't Hooft observables
\ie
W_m(\tilde{\mathcal M}^{(d-p-2)}) = \exp\left[ i \sum_{\text{dual }(d-p-2)\text{-cell}\in \tilde{\mathcal M}^{(d-p-2)}} \tilde a^{(d-p-2)} \right]~,
\fe
where $\tilde{\mathcal M}^{(d-p-2)}$ is a codimension-$(p+2)$ submanifold along the dual $(d-p-2)$-cells of the lattice.

\subsubsection{Long-distance limit}

In the continuum limit, the modified Villain model \eqref{U1-pform-modVill-action} becomes a gapless continuum $p$-form gauge theory
\ie\label{highforg}
\frac{1}{2g^2}\int d^dx~(da^{(p)})^2~.
\fe
This can be derived, as above, by choosing a convenient gauge where most of the integer-valued fields vanish and then redefining the real lattice variables appropriately.\footnote{The continuum theory can also have additional $\theta$-parameters associated with various characteristic classes of the gauge field.  Our lattice formulation leads to the term ${\theta\over 2\pi} da^{(p)}$ for $p=d-1$, but not to the other $\theta$-parameters. For example, see \cite{Sulejmanpasic:2019ytl,Anosova:2019quw} for a discussion on the $\theta$-parameter in the modified Villain version of the ordinary 3+1d $U(1)$ gauge theory.}

An important question is whether the lattice gauge theory \eqref{U1-pform-action}, or equivalently its Villain version \eqref{U1-pform-Vill-action}, flow at long distances to the same gapless theory \eqref{highforg}. Unlike the modified Villain model, these two lattice models have only the electric symmetry, but no magnetic symmetry.  So without fine-tuning, the long-distance theory is generically deformed by the 't Hooft operators. For the deformation to be possible, the 't Hooft operators have to be local, point-like operators. This is the case only for $p=d-2$. This is obvious in its dual version where the dual field is a scalar and the monopole operator gives it a mass.  This implies that without fine-tuning a $d$-dimensional $p$-form lattice gauge theory can flow to a gapless $p$-form gauge theory at long distance unless $p=d-2$, in which case, the theory is generically gapped at long distance.  This is the famous Polyakov mechanism \cite{POLYAKOV1977429}.

We conclude that for $p=d-2$, where the standard $U(1)$ lattice gauge theory is gapped, the modification of the lattice gauge theory \eqref{U1-pform-modVill-action} keeps it massless.

\subsection{$\mathbb Z_N$ gauge theory}\label{ZNgauget}

We now describe a $d$-dimensional Villain $\mathbb{Z}_N$ $p$-form gauge theory \cite{PhysRevD.19.3698,Ukawa:1979yv}. On each $p$-cell, there is an integer field $m^{(p)}$ and on each $(p+1)$-cell, there is an integer field $n^{(p+1)}$. The action is
\ie\label{ZNpform}
\frac{\beta(2\pi)^2}{2N^2}\sum_{(p+1)\text{-cell}}(\Delta m^{(p)}-Nn^{(p+1)})^2~,
\fe
with the integer gauge symmetry
\ie
&m^{(p)}\sim m^{(p)}+\Delta \ell^{(p-1)}+Nk^{(p)}~,
\\
&n^{(p+1)}\sim n^{(p+1)}+\Delta k^{(p)}~.
\fe
The theory has an electric $\mathbb{Z}_N$ $p$-form global symmetry \cite{Gaiotto:2014kfa}, which shifts $m^{(p)}$ by a flat integer $p$-form field.

In the limit $\beta\rightarrow\infty$, the field strength obeys $\Delta m=0$ mod $N$ \cite{PhysRevD.19.3682,Banks:1979fi}, and we can replace the action by
\ie\label{eq:ZNtopoaction}
\frac{2\pi i}{N}\sum_{p\text{-cell}}m^{(p)}\Delta {\tilde n}^{(d-p-1)}~,
\fe
where ${\tilde n}^{(d-p-1)}$ is an integer-valued field with the integer gauge symmetry
\ie
&\tilde n^{(d-p-1)}\sim \tilde n^{(d-p-1)}+\Delta\tilde k^{(d-p-2)}+N\tilde q^{(d-p-1)}~.
\fe
This describes a \textit{topological $\mathbb{Z}_N$ lattice gauge theory} \cite{Dijkgraaf:1989pz,Kapustin:2014gua}. The action \eqref{eq:ZNtopoaction} is similar to the one in \cite{Kapustin:2014gua} except that the fields there are $\mathbb{Z}_N$ variables while here we use $\mathbb{Z}$ variables with $N\mathbb{Z}$ gauge symmetry.

\subsubsection{Duality}

As in Appendix \ref{sec:ZNclock}, we can dualize the $\mathbb{Z}_N$ $p$-form gauge theory \eqref{ZNpform} by dualizing the integer field $n^{(p+1)}$  to an integer field $\tilde n^{(d-p-1)}$:
\ie
\frac{1}{2\beta}\sum_{\text{dual $(d-p-1)$-cell}}(\tilde n^{(d-p-1)})^2+\frac{2\pi i}{N}\sum_{p\text{-cell}}m^{(p)}\Delta \tilde n^{(d-p-1)} ~.
\fe
For $p\leq d-1$, we can introduce new gauge symmetries together with Stueckelberg fields, and write the action as
\ie
\frac{1}{2\beta}\sum_{\text{dual $(d-p-1)$-cell}}(\Delta\tilde m^{(d-p-2)}-N\hat n^{(d-p-1)}-\tilde n^{(d-p-1)})^2+\frac{2\pi i}{N}\sum_{p\text{-cell}}m^{(p)}\Delta \tilde n^{(d-p-1)} ~.
\fe
with the integer gauge symmetry
\ie
&\tilde m^{(d-p-2)}\sim \tilde m^{(d-p-2)}+\Delta\tilde\ell^{(d-p-3)}+\tilde k^{(d-p-2)}~,
\\
&\hat n^{(d-p-1)}\sim\hat n^{(d-p-1)}-\tilde q^{(d-p-1)}~,
\\
&\tilde n^{(d-p-1)}\sim \tilde n^{(d-p-1)}+\Delta\tilde k^{(d-p-2)}+N\tilde q^{(d-p-1)}~,
\\
&m^{(p)}\sim m^{(p)}+\Delta\ell^{(p-1)}+Nk^{(p)}~.
\fe
The duality maps a $p$-form gauge theory with coefficient $\frac{2\pi^2\beta}{N^2}$ to a $(d-p-2)$-form gauge theory with coefficient $\frac{1}{2\beta}$ that couples to a topological $\mathbb{Z}_N$ $(d-p-1)$-form gauge theory.
For $d=2$ and $p=0$, this reduces to the Kramers-Wannier duality of the $\bZ_N$ clock model reviewed in Appendix \ref{KWduality}.
The duality of the $d=3$ and $p=1$ system is the famous duality of the 3d clock model \cite{Wegner:1984qt,Savit:1979ny,PhysRevD.19.3715} and for $d=4$ and $p=1$ it is the famous self-duality of \cite{Wegner:1984qt,Savit:1979ny,PhysRevD.17.2637,PhysRevD.19.3715,PhysRevD.19.3698,Ukawa:1979yv}.

\subsubsection{Real $BF$-action and the continuum limit}

This theory can be described using several different actions.  Here we describe some actions using real fields that are similar to various continuum actions.

We start with the integer $BF$-action \eqref{eq:ZNtopoaction} and replace the integer-valued gauge fields $m^{(p)}$ and $\tilde n^{(d-p-1)}$ with real-valued gauge fields $a^{(p)}$ and $\tilde b^{(d-p-1)}$. We constrain these real-valued fields to integer values by adding integer-valued fields $\tilde m^{(d-p)}$ and $n^{(p+1)}$. Furthermore, since the gauge fields $a^{(p)}$ and $\tilde b^{(d-p-1)}$ have real-valued gauge symmetries instead of integer-valued gauge symmetries, we introduce Stueckelberg fields $\phi^{(p-1)}$ and $\tilde\phi^{(d-p-2)}$ for the gauge symmetries.  We end up with the action
\ie\label{eq:ZN_topo_action_real}
&\frac{ iN}{2\pi}\sum_{p\text{-cell}}a^{(p)}\left(\Delta \tilde b^{(d-p-1)} -2\pi  \tilde m^{(d-p)}\right)+i(-1)^pN\sum_{(p+1)\text{-cell}}n^{(p+1)}\tilde b^{(d-p-1)}
\\
&\quad -i(-1)^p\sum_{(p+1)\text{-cell}} n^{(p+1)}\Delta\tilde\phi^{(d-p-2)} +i\sum_{p\text{-cell}} \Delta\phi^{(p-1)} \tilde m^{(d-p)}~.
\fe
We will refer to this presentation of the model as the real $BF$-action, which uses both real and integer fields.

As a check, summing over $\tilde m^{(d-p)}$ and $n^{(p+1)}$ constrains
\ie
a^{(p)}-\frac{1}{N}\Delta \phi^{(p-1)}=\frac{2\pi}{N}m^{(p)}~,\qquad\tilde b^{(d-p-1)}-\frac{1}{N}\Delta\tilde \phi^{(d-p-2)}=\frac{2\pi}{N}\tilde n^{(d-p-1)}~,
\fe
where $m^{(p)}$ and $\tilde n^{(d-p-1)}$ are integer-valued fields. Substituting them into \eqref{eq:ZN_topo_action_real}, we recover the action \eqref{eq:ZNtopoaction}.

The action \eqref{eq:ZN_topo_action_real} has the gauge symmetry
\ie\label{eq:ZNgaugesym}
&a^{(p)}\sim a^{(p)}+\Delta\alpha^{(p-1)}+2\pi k^{(p)}~,
\\
&\tilde b^{(d-p-1)}\sim \tilde b^{(d-p-1)}+\Delta\tilde \beta^{(d-p-2)}+2\pi\tilde q^{(d-p-1)}~,
\\
&n^{(p+1)}\sim n^{(p+1)}+\Delta k^{(p)}~,
\\
&\tilde m^{(d-p)}\sim \tilde m^{(d-p)}+\Delta\tilde q^{(d-p-1)}~,
\\
&\phi^{(p-1)}\sim \phi^{(p-1)}+\Delta \gamma^{p-2}+N\alpha^{(p-1)}+2\pi k_\phi^{(p-1)}~,
\\
&\tilde\phi^{(d-p-2)}\sim\tilde\phi^{(d-p-2)}+\Delta\tilde\gamma^{(d-p-3)}+N\tilde \beta^{(d-p-2)}+2\pi \tilde q_{\tilde\phi}^{(d-p-2)}~,
\fe
where $\alpha^{(p-1)},\tilde\beta^{(d-p-2)},\gamma^{(p-2)},\tilde\gamma^{(d-p-3)}$ are real-valued  and $k_\phi^{(p-1)},\tilde q_{\tilde\phi}^{(d-p-2)}$ are integer-valued.

Another action is obtained by replacing $\tilde m^{(d-p)}$ by a real-valued field $\tilde F^{(d-p)}+\Delta\tilde b^{(d-p-1)}$ and adding an integer-valued field $n^{(p)}$ to constrain it. This leads to the action
\ie\label{eq:ZNHiggs}
&\frac{i}{2\pi}\sum_{p\text{-cell}} (\Delta\phi^{(p-1)} -Na^{(p)}-2\pi n^{(p)})\tilde F^{(d-p)} +i(-1)^p\sum_{(p+1)\text{-cell}}(\Delta n^{(p)}+Nn^{(p+1)})\tilde b^{(d-p-1)}
\\
&\quad -i(-1)^p\sum_{(p+1)\text{-cell}} n^{(p+1)}\Delta\tilde\phi^{(d-p-2)} ~.
\fe
These fields have the same gauge symmetries as in \eqref{eq:ZNgaugesym}. In addition, the gauge symmetries also act on $n^{(p)}$
\ie
n^{(p)}\sim n^{(p)}+\Delta k_\phi^{(p-1)}-Nk^{(p)}~.
\fe
We can interpret the action \eqref{eq:ZNHiggs} as Higgsing the $U(1)$ gauge theory of $a^{(p)}$ to a $\mathbb{Z}_N$ theory using the fields $\phi^{(p-1)}$ of charge $N$.

Alternatively, we can integrate out $\phi^{(p-1)},\tilde\phi^{(d-p-2)}$ which constrain $n^{(p+1)},\tilde m^{(d-p)}$ to be flat gauge fields. Using the gauge symmetry of $k^{(p)},\tilde q^{(d-p-1)}$, we can gauge fix $n^{(p+1)},\tilde m^{(d-p)}$ to be zero almost everywhere except at a few cells that capture the holonomy. The residual gauge symmetry shifts $a^{(p)}$ and $\tilde b^{(d-p-1)}$ by $2\pi$ multiples of flat integer gauge fields. Let us define two new fields $\bar a^{(p)},\bar{\tilde b}^{(d-p-1)}$ such that
\ie
&\Delta \bar a^{(p)}=\Delta a^{(p)}-2\pi n^{(p+1)}~,
\\
&\Delta \bar{\tilde b}^{(d-p-1)}=\Delta \tilde b^{(d-p-1)}-2\pi \tilde m^{(d-p)}~,
\fe
and $\bar a^{(p)}=a^{(p)}$, $\bar{\tilde b}^{(d-p-1)}=\tilde b^{(d-p-1)}$ almost everywhere. Although $a^{(p)},\tilde b^{(d-p-1)}$ are single-valued fields, $\bar a^{(p)},\bar {\tilde b}^{(d-p-1)}$ can have nontrivial transition functions. In terms of the new variables, the Euclidean action is
\ie\label{ZNgaugefixaction}
\frac{ iN}{2\pi}\sum_{p\text{-cell}}\bar a^{(p)}\Delta \bar{\tilde b}^{(d-p-1)}+i(-1)^pN\sum_{(p+1)\text{-cell}} n^{(p+1)}\bar{\tilde b}^{(d-p-1)}~,
\fe
where $n^{(p+1)}$ vanishes almost everywhere except at a few $(p+1)$-cells, which encode the information in the transition function of $\bar a^{(p+1)}$.  For $d=1$ and $p=0$, the action \eqref{ZNgaugefixaction} reduces to the quantum mechanics action  \eqref{gaugecom}.

The real $BF$-action is closely related to the continuum field theory limit.
In this gauge choice,  the continuum limit is
\ie\label{ZNcontcc}
\frac{iN}{2\pi}\int\,  a^{(p)}d \tilde b^{(d-p-1)}~,
\fe
where we dropped the bars on $a^{(p)}$ and $\tilde b^{(d-p-1)}$ and rescaled them by appropriate powers of the lattice spacing $a$. We also omitted here the terms that depend on the transition functions of $a^{(p)}$ and $\tilde b^{(d-p-1)}$. As in \eqref{eq:pqdotactionc}, these terms are actually essential in order to make \eqref{ZNcontcc} globally well defined. Here $a^{(p)}$ is a $U(1)$ $p$-form gauge field and $\tilde b^{(d-p-1)}$ is a $U(1)$ $(d-p-1)$-form gauge field. This is the known continuum action of the $\mathbb{Z}_N$ $p$-form gauge theory \cite{Maldacena:2001ss,Banks:2010zn,Kapustin:2014gua}.

\subsubsection{Relation to the toric code}

We now review the well-known fact that the low-energy limit of the $\mathbb{Z}_N$ toric code \cite{Kitaev:1997wr} is described by the topological $\mathbb{Z}_N$ lattice gauge theory \cite{Dijkgraaf:1989pz,Kapustin:2014gua}, which in turn is given by the continuum $\bZ_N$ gauge theory.

Consider the $\mathbb{Z}_N$ toric code  on a 2d periodic square lattice. On each link, there is a $\mathbb{Z}_N$ variable $U$ and its conjugate variable $V$. They obey $UV =e^{2\pi i/N}VU $ and $U^N=V^N=1$. The Hamiltonian consists of two commuting terms $G$ and $L$:
\ie
H_{\text{toric}}=-\beta_1\sum_{\text{site}}G -\beta_2\sum_{\text{plaq}}L+c.c.~,
\fe
where $G$ is an oriented product of $V$ and $V^\dagger$ around a site and $L$ is an oriented product of $U$ and $U^\dagger$ around a plaquette.

The ground states satisfy $G=L=1$ for all sites and plaquettes, while the excited states violate some of these conditions. It is common to refer to the dynamical excitations that violate only $G=1$ at a site as the electrically-charged excitations and those that violate only $L=1$ at a plaquette as the magnetically-charged excitations.

The toric code has a large non-relativistic electric and magnetic $\mathbb{Z}_N$ one-form symmetry (in the sense of \cite{Seiberg:2019vrp}).
The symmetries are generated respectively by the closed loop operator $W_e$ made of $V$ and $V^\dagger$, and the closed loop operator $W_m$ made of $U$ and $U^\dagger$.  Unlike the relativistic one-form symmetry of  \cite{Gaiotto:2014kfa}, these symmetry operators are not topological, i.e., they are not invariant under small deformations.

In the $\beta_1,\beta_2\rightarrow\infty$ limit, the Hilbert space is restricted to the ground states, which satisfy $G=L=1$ for all sites and plaquettes. In the restricted Hilbert space, there are no electrically-charged or magnetically-charged excitations. So, the closed loop operators $W_e$ and $W_m$ are topological, and they generate a relativistic electric and magnetic $\mathbb{Z}_N$ symmetry, respectively.

Consider the  toric code in the $\beta_1,\beta_2\rightarrow\infty$  limit in the Lagrangian formalism on a 3d Euclidean lattice. For each spatial link along the $i=x,y$ direction, we introduce an integer field $m_{i}$ for the $\mathbb{Z}_N$ variable $U=\exp(\frac{2\pi i }{N}m_i )$, and an integer field $\tilde n_{j}$ for the conjugate $\mathbb Z_N$ variable $V=\exp(\frac{2\pi i}{N}\epsilon^{ij}\tilde n_j)$. The field $\tilde n_j$ naturally lives on the dual links along the $j$ direction.

To impose the constraints $G=L=1$, we  introduce two integer-valued Lagrange multiplier fields.
On each $\tau$-link, we introduce an integer field $m_\tau$ to impose $G=1$, or equivalently $\epsilon^{ij}\Delta_i\tilde n_j=0$ mod $N$. On each dual $\tau$-link (or equivalently each $xy$-plaquette), we introduce an integer field $\tilde n_\tau$ to impose $L=1$, or equivalently $\epsilon^{ij}\Delta_i m_j=0$ mod $N$. In terms of these integer fields, the Euclidean action of the system is precisely the topological $\mathbb{Z}_N$ lattice gauge theory \eqref{eq:ZNtopoaction} with $d=3$ and $p=1$.

\bibliographystyle{JHEP}
\bibliography{Villain}

\end{document}